\documentclass{ieeeaccess}

\usepackage{cite}
\usepackage{amsmath,amssymb,amsfonts}
\usepackage{algorithmic}
\usepackage{algorithm}
\usepackage{graphicx}
\usepackage{textcomp}
\usepackage{subcaption}

\usepackage{booktabs}
\usepackage{multirow}

\usepackage[utf8]{inputenc}

\usepackage{url}

\def\BibTeX{{\rm B\kern-.05em{\sc i\kern-.025em b}\kern-.08em
    T\kern-.1667em\lower.7ex\hbox{E}\kern-.125emX}}
\begin{document}

\history{Date of publication xxxx 00, 0000, date of current version xxxx 00, 0000.}
\doi{10.1109/ACCESS.2017.DOI}

\title{Compressed Data Structures for Binary Relations in Practice}
\author{\uppercase{Carlos Quijada-Fuentes}\authorrefmark{1},  \uppercase{Miguel R. Penabad}\authorrefmark{2}, 
 \uppercase{Susana Ladra}\authorrefmark{2}, and \uppercase{Gilberto Gutiérrez}\authorrefmark{1},}
\address[1]{Universidad del Bío-Bío, Facultad de Ciencias Empresariales, 3800708, Chillán, Chile (e-mail: ggutierr@ubiobio.cl)}
\address[2]{Universidade da Coruña, Centro de investigación CITIC, Facultade de Informática, 15071, A Coruña, Spain (e-mail: \{miguel.penabad,susana.ladra\}@udc.es)}
\tfootnote{This research has received funding from the European Union's Horizon 2020 research and innovation programme under the Marie Sklodowska-Curie [grant agreement No 690941]; from the Ministerio de Ciencia, Innovaci{\'o}n y Universidades (PGE and ERDF) [grant numbers TIN2016-77158-C4-3-R; TIN2016-78011-C4-1-R; RTC-2017-5908-7]; Consellería de Economía e Industria of the Xunta de Galicia through the GAIN (Axencia Galega de Innovación), co-funded with ERDF [grant number IN852A 2018/14]; from
Xunta de Galicia (co-funded with ERDF) [grant numbers ED431C 2017/58; ED431G/01]; and from University of Bío-Bío [grant numbers 192119 2/R  and 195119 GI/VC].} 
%Proyecto de investigación, 192119 2/R  and  grupo de investigación ALBA: Algoritmos y Bases de Datos, 195119 GI/VC, ambos financiados por la Universidad del Bío-Bío.

\markboth
{Quijada-Fuentes \headeretal: Compressed Data Structures for Binary Relations in Practice}
{Quijada-Fuentes \headeretal: Compressed Data Structures for Binary Relations in Practice}

\corresp{Corresponding author: Carlos Quijada-Fuentes (e-mail: caquijad@egresados.ubiobio.cl).}

\begin{abstract}
Binary relations are commonly used in Computer Science for modeling data. In addition to classical representations using matrices or lists, some compressed data structures have recently been proposed to represent binary relations in compact space, such as the $k^2$-tree and the Binary Relation Wavelet Tree (BRWT).
Knowing their storage needs, supported operations and time performance is key for enabling an appropriate choice of data representation given a domain or application, its data distribution and typical operations that are computed over the data.

In this work, we present an empirical comparison among several compressed representations for binary relations. We analyze their space usage and the speed of their operations using different (synthetic and real) data distributions. We include both neighborhood and set operations, also proposing algorithms for set operations for the BRWT, which were not presented before in the literature. We conclude that there is not a clear choice that outperforms the rest, but we give some recommendations of usage of each compact representation depending on the data distribution and types of operations performed over the data. We also include a scalability study of the data representations. 
\end{abstract}


\titlepgskip=-15pt

\maketitle

\section{Introduction}
 Let $A$ and $B$ be two sets of objects. A binary relation $R$ is defined 
as a subset of the Cartesian product $A \times B$, where for each element $(a,b) \in R$, 
we say that $a$ is related to $b$ and denote this as $aRb$.
In the areas of Mathematics and Computer Science, binary relations constitute a fundamental
conceptual and methodological tool \cite{DOUGHERTY2006228} used for representing properties or relationships
among objects in a simple and intelligent way \cite{Ah2013}. 
In Computer Science, binary relations can be modeled by using data structures such as graphs,
trees, inverted indices, or discrete grids \cite{BBN12,BCPdBNis17}. By using binary relations,
it is possible to model complex problems. For instance, connections among pages of a particular
Web site, or even among all pages in the World Wide Web (WWW) \cite{Broder2000, BV2004, Papadimitriou2010, BLN14}; 
other fields are automated recommendation systems, where the users (customers) are related to
purchased products \cite{ABP2005,HFY2015,Nav16}. 

% NO SÉ SI AQUÍ ES EL MEJOR SITIO PARA DEFINIR LAS OPERACIONES...
For these relations we can name a number of \emph{Neighborhood} queries, such as those defined in \cite{yao2004semantics}. Consider a relation $R \subseteq A \times B$,
with a total ordering $\leq_A$ in $A$ and a total ordering $\leq_B$ in $B$. Then, we define the following operations:
\begin{itemize}
    \item $isRelated(x,y) = true \mbox{ if } xRy, ~false \mbox{ otherwise}$.
    \item $successors(x) = \{y \in B| xRy\}$
    \item $predecessors(y) = \{x \in A | xRy\}$
    % reviewer 1 Q 5: change y2 to y_2
    \item $rangeNeighborhood(x_1,y_1,x_2,y_2) = \{(x,y) | xRy,$ $x_1 \leq_A x\leq_A x_2 \wedge y_1 \leq_B y \leq_B y_2 \} $
\end{itemize}

Binary relations have been traditionally stored using either adjacency matrices or adjacency lists.
However, and due to the growth on the size that the sets of binary relations currently generated
are experiencing (for example, a graph of the whole WWW), it is convenient to store these sets
using compact data structures. The goal of doing so is to reduce the storage needs (either RAM or disk),
but maintaining the capacity of processing the data directly in their compressed form.
Reducing the storage size may have the advantage of diminishing, even removing, the need for I/O operations.

One of the most widely known compact data structures used to store binary relations 
are the $k^2$-tree \cite{BLN14} and the Binary Relation Wavelet Tree (BRWT) \cite{BCN13},
which is based on a Wavelet Tree \cite{Grossi:2003:HET:644108.644250}. In general, these
structures support only some basic operations. For example, the $k^2$-tree was initially
proposed to represent Web graphs, so it implemented operations such as $isRelated$ (which tests whether page
$X$ links to page $Y$, called $access$ in the original paper), finding successor or predecessor neighbors,
%(which were direct and reverse neighbors in the original paper), 
and range neighborhood queries. 
More recently, in \cite{QUIJADAFUENTES201976}, the operations over $k^2$-trees were extended to include set
operations, that is, union, intersection, or difference, among others. In the case of BRWTs, there is 
a number of operations defined over them, such as 
primitive operations obtaining the labels associated to a given object, or the range of objects
associated to a given label. However, no set operations are defined over BRWTs.

When deciding which compact data structure to choose for representing binary relations in the context of 
a given domain or application, it is very convenient to know in advance the adequacy of the available
data structures, in terms of storage needs, supported operations, and time performance. %, usually measured
%as the amount of time taken to perform the operations. 
The decision can also consider the frequency of each kind of operation. For example, one application might make an intensive use of union and 
intersection operations, but rarely searches for predecessor neighbors or performs range neighborhood 
queries.

In this work, we present a comparison of three compact data structures that can be used to
represent binary relations: $k^2$-tree, $k^2$-tree1 and BRWT (Binary Relation Wavelet Tree).
The comparison considers the same operations for all evaluated data structures. Basically,
they are set operations (union, intersection, difference, and symmetric difference)
and primitive neighborhood operations (isRelated, successors, predecessors, and range neighborhood queries). 
The goal of this comparison is to facilitate the choice of the most accurate data structure for
a particular application or domain.
As an additional alternative to compact data structures, we also include in our comparison the 
representation of the binary relations using compressed adjacency lists (using QMX and Rice-runs encoders).

Another contribution of the current work is the design and implementation of all of the algorithms needed to perform
the set operations and the neighborhood queries over binary relations represented with BRWT.
Like the operations we use for $k^2$-trees and $k^2$-tree1s, these algorithms operate directly 
over the compact data structures, without decompressing them.

The rest of this paper is organized as follows.
Section~\ref{previousWork} shows a review of the compact data structures considered in the comparison.
Section~\ref{BWRT} describes the algorithms for performing set operations over BRWTs.
Section~\ref{sec:experiments} shows our empirical evaluation.
Finally, the last section offers the overall discussion of the results and some conclusions of this work.

\section{Previous work}
\label{previousWork}
In this section, we describe the compact data structures and encoders that will be used in our comparison.

\subsection{$k^2$-tree}
A $k^2$-tree \cite{BLN14} is a succinct data structure originally designed to represent Web graphs,
but it is able to represent any binary relation. 
 
A $k^2$-tree for a binary relation represented by a matrix of size $n\times n$ is built as follows\footnote{If the matrix is not squared or $n$ is not a power of $k$, it is conceptually extended to the right and to the bottom with 0s, rounding the size up to the next power of $k$. This does not cause
a significant overhead because the $k^2$-tree can handle large areas of 0s efficiently.}: 
the root node is associated to the whole matrix, which is divided into $k^2$ submatrices ($k$ rows
by $k$ columns). For each of these submatrices, a child is added to the root node. We store a 0 in
the node if all cells in the submatrix are 0s, or a 1 if any cell contains a 1. We then proceed 
recursively on all children associated to a 1. 
% reviewer 2 Q2
%The children associated with a 0 are no further decomposed,
The recursion stops when the algorithm processes either an individual cell or a submatrix of 0s,
so the resulting tree is not balanced.

By design, $k^2$-trees perform very well when the matrix has a relatively low number of 1s that are
clustered together, because large areas of 0s are represented by a single 0 bit in the $k^2$-tree.
However, each 1 in the matrix can use more than one bit in the $k^2$-tree, so its behavior worsens 
if the number of 1s increases. To avoid this problem, a variation of the $k^2$-tree, which we denote 
$k^2$-tree1 in this work, was designed in \cite{BAB+13}. Basically, it represents a uniform
submatrix (either full of ones or zeroes) by a single 0 (with an additional bitmap to decide whether
it is full of ones or zeroes) and mixed submatrices with ones and zeroes by a 1. The recursion
proceeds only for these mixed areas.

Being succinct data structures, the described conceptual trees are not stored, but only the bitmaps
of their nodes. Navigation operations over $k^2$-trees and $k^2$-tree1s are described in \cite{BLN14} and \cite{BAB+13},
respectively. Set operations for both, including the pseudocode for the algorithms as well as empirical results on 
their performance, are described in \cite{QUIJADAFUENTES201976}.

\subsection{BRWT} \label{subsec-brwt}
\begin{figure*}[t]
    \centering
    \begin{subfigure}[A binary relation matrix]{0.3\textwidth}{\includegraphics[width=0.98\textwidth]{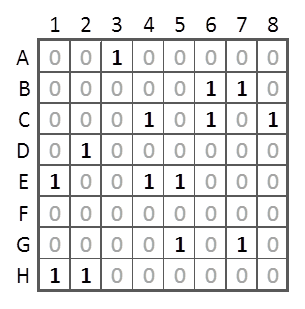}}    
    \caption{A binary relation matrix}
    \end{subfigure}
    \begin{subfigure}[Its BRWT representation]{0.6\textwidth}{\includegraphics[width=0.98\textwidth]{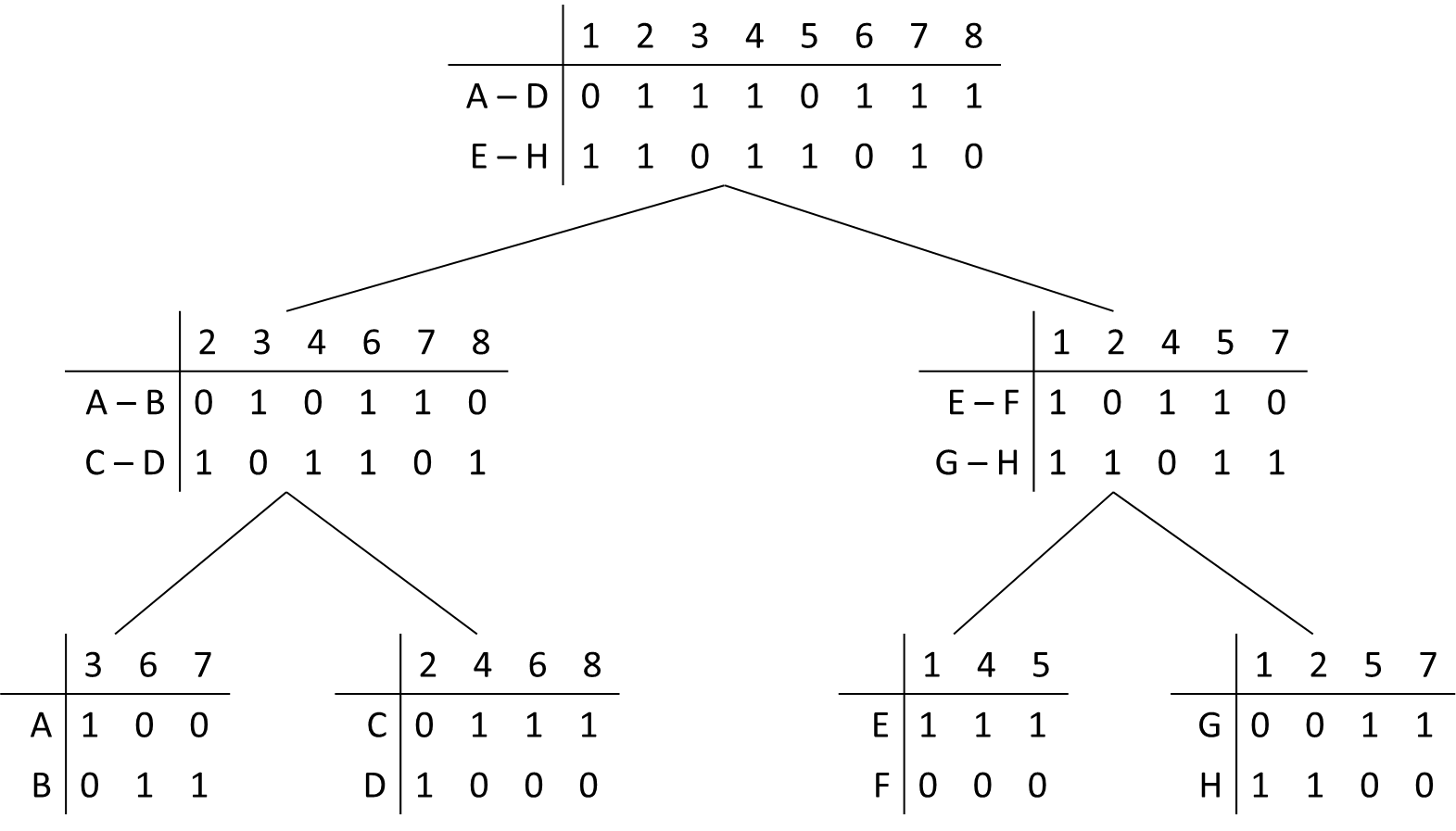}}
    \caption{Its BRWT representation}
    \end{subfigure}
    \caption{A BRWT example (based on \protect\cite{BCN13}).}
    \label{fig:brwt}
\end{figure*}

A Binary Relation Wavelet Tree, or BRWT  \cite{BCN13}, is a special type of wavelet tree \cite{Grossi:WT} 
specifically designed to represent binary relations. An example of the conceptual tree built for a given 
binary matrix is show in Figure \ref{fig:brwt}. Each node contains two bitmaps, which correspond to two 
submatrices: the top and bottom halves of the original matrix. A bitmap position is set to 0 if all values 
in this column of the submatrix are 0s (as in column 1 for the A-D bitmap in the root node) and it is set 
to 1 if any cell in this column has a 1 (column 2 at the same bitmap).

The left and right subtrees are built recursively considering the bits set to 1 at the top/bottom
bitmaps. For example, column 1 does not appear in the left subtree of the root node. Note also that, 
like column 2, a column can propagate to both left and right subtrees. This fact makes the BRWT
different from the original wavelet trees, because the bits of a given level may be more than $n$, the
number of objects.
Interested readers can find in  \cite{BCN13} more information about the operations supported 
by BRWTs and some bounds on their complexity. We shall provide in the next section 
detailed information, as well as 
the pseudocode on which our implementation is based, of the set operations 
%and neighborhood algorithms
for binary relations implemented in BRWTs.

\subsection{Compressed adjacency lists} \label{subsec-qmx-rice}
A very naive and (generally) space-consuming representation of a binary relation is an adjacency list.
Although a valid representation, it is not suitable if the relation is big and does not
fit into main memory. To avoid this problem, the adjacency lists can be compressed.

There are multiple techniques for compressing lists of integers, such as 
%authors have tested in \cite{QUIJADAFUENTES201976} two different techniques: 
QMX and Rice-runs.
QMX \cite{Tro14} is a compression algorithm that combines word-alignment, SIMD instructions, and 
run-length encoding. It also includes a SIMD-aware intersection algorithm \cite{LBK16}. 
Rice-runs combines the well-known Rice coding \cite{WMB99} with run-length compression \cite{CM10,TS10}. 
QMX performs really well for long adjacency lists, where SIMD instructions can be exploited.  
Rice-runs is specially suitable when the lists contain large sequences of consecutive 1s in the input 
relation matrix, due to the use of run-length compression, boosting both compression and intersection speed. 

These techniques can be applied to compress any input. The authors have already used them to compress binary relations,
using their own implementation in \cite{QUIJADAFUENTES201976}. We must note, however, that these
are \emph{not} compact data structures. They are only compression schemes, and the lists must be decompressed
before they can be used to efficiently perform the requested operations on the uncompressed data.

\section{Set Operations over BRWT}
\label{BWRT}
We now describe the algorithms for computing union, intersection, difference, and symmetric difference 
of binary relations represented using BRWTs. 

Like the $k^2$-tree, BRWT is a hierarchical structure. Thus, the approach for the algorithms described in
\cite{QUIJADAFUENTES201976, BrisaboaBGLPT15} to implement set operations over $k^2$-trees can be applied 
here. Of course, the differences and specific properties of the BRWT must be taken into account.

The algorithms use essentially a breadth-first traversal of the trees that represent the input relations
for the union, and a depth-first traversal for the remaining operations. The algorithm for the union is
presented in Subsection~\ref{subsec-anchura}, and the algorithm for the intersection is described in 
Subsection ~\ref{subsec-profundidad}. The remaining operations, difference and symmetric difference,
use an algorithm very similar to the intersection, so they are briefly described in the same subsection.

For this section, given a node $b$ of the BRWT, $b_l$ and $b_r$ represent the bitmaps associated to $b$.
For instance, considering $b$ the root node of the BRWT at Figure~\ref{fig:brwt},  $b_l$ is $01110111$, and $b_r$ is $11011010$.

\subsection{Breadth-first traversal algorithm (Union)} \label{subsec-anchura}

Given the properties of the union, if we are processing two nodes (of the two input BRWT), it is
possible to obtain the result without accessing the children of these nodes. That is, if a bit 
representing a column in one of the nodes is $1$, the output is $1$ regardless of the value of the bit
in the other node, and if both bits are $0$, the output is $0$. This enables the use of a breadth-first
traversal over the BRWTs to compute the result of the union operation.

The traversal is performed by doing a synchronized sequential scan of the two input BRWT bitmaps $A$ and $B$.
Note that this synchronization must take into account that there can exist a column that is
defined in one of the two nodes being processed, but not in the other. 
Algorithm~\ref{algorithm_union_wt} uses queues (as usual for breadth-first traversal of any tree).
In this case, each element of the queue is a pair of flags $\langle f_A, f_B \rangle$ (indicating if
the current column is defined in inputs $A$ and $B$, respectively), which is used
to determine the output bit, and whether we should enqueue a new pair to process the current column
in the child nodes (at the next level of the conceptual tree). 

The algorithm actually uses two queues: $Q_l$ and $Q_r$. The reason behind their use is the way the
bitmaps are stored: for each node $b$, we store the $b_l$ followed by the $b_r$. Then, $Q_l$ is used
to manage the breadth-first traversal of the BRWT, while $Q_r$ is only used to process the $b_r$ part
of each node (we can see in the algorithm that when an element is dequeued from $Q_l$, it is enqueued
in $Q_r$, but the enqueuing needed to process lower levels of the conceptual tree is always done in
$Q_l$).

%%%%%%%%%%%%%%%%%%%%%%%%%%%%%%%%%%%%%%%%%%%%%%%%%%%%%%%%%%%%
%%%%%%%%%% El Algoritmo Unión %%%%%%%%%%%%%%%%%%%%%%%%%%%%%%
%%%%%%%%%%%%%%%%%%%%%%%%%%%%%%%%%%%%%%%%%%%%%%%%%%%%%%%%%%%%
\begin{algorithm}[t]
	\caption{\textbf{UnionBRWT}($A,B$)}
	\scriptsize
	\algsetup{linenosize=\scriptsize}
	\begin{algorithmic}[1]
		\STATE $useQ_l \leftarrow true$ \label{useQ_lvar}
		\STATE $pA \leftarrow 0, pB \leftarrow 0$
		\STATE $bA \leftarrow 0, bB \leftarrow 0$
		\FOR{$i\leftarrow 0\dots numOfObjects$}
		\STATE $Q_l.Insert(\langle 1, 1\rangle)$
		\ENDFOR
		\STATE $Q_l.Insert(\langle 0, 0\rangle)$
		\WHILE{$pA < |A| \vee pB < |B|$ }
		\IF{$useQ_l$} \label{useQ_luse1}
		\STATE $\langle f_A, f_B \rangle \leftarrow Q_l.Remove()$
		\ENDIF
		\STATE $children \leftarrow 0$
		\WHILE{$((useQ_l \wedge (f_A \vee f_B)) \vee (!useQ_l \wedge !Q_rIsEmpty()))$ }
		\IF{$useQ_l$}
		\STATE $Q_r.Insert(\langle f_A, f_B \rangle)$
		\ELSE
		\STATE $\langle f_A, f_B \rangle \leftarrow Q_r.Remove()$
		\ENDIF
		\STATE $bA \leftarrow f_A \wedge A[pA]$
		\STATE $bB \leftarrow f_B \wedge B[pB]$
		\IF{$(bA \vee bB) \wedge (!isLeaf(pA) \vee !isLeaf(pB))$}
		\STATE $Q_l.Insert(\langle bA, bB \rangle)$
		\STATE $children \leftarrow children + 1$
		\ENDIF
		\STATE $R[posR] \leftarrow bA \vee bB$ \label{wt_union_RposR}
		\STATE $posR \leftarrow posR + 1$
		\IF{$f_A$}
		\STATE $pA \leftarrow pA + 1$ \label{wt_union_pA++}
		\ENDIF
		\IF{$f_B$}
		\STATE $pB \leftarrow pB + 1$ \label{wt_union_pB++}
		\ENDIF
		\IF{$useQ_l$} \label{useQ_luse2}
		\STATE $\langle f_A, f_B \rangle \leftarrow Q_l.Remove()$
		\ENDIF
		\ENDWHILE
		\STATE $useQ_l \leftarrow (\sim useQ_l)$
		\IF{$(!isLeaf(pA) \vee !isLeaf(pB)) \wedge children > 0$}
		\STATE $Q_l.Insert(\langle 0,0 \rangle)$
		\ENDIF
		\ENDWHILE
		\RETURN $R$
	\end{algorithmic}
	\label{algorithm_union_wt}
\end{algorithm}

\subsection{Depth-first traversal algorithms (intersection, difference, and symmetric difference)} \label{subsec-profundidad}

The algorithms for the intersection, difference and symmetric difference use a depth-first traversal
of the input BRWTs, because the output bit for a column in a node depends on the values of the same
column in the descendants of this node, down to the leaves. The algorithms for the three operations are
very similar, in fact the navigation scheme is exactly the same for all of them. The only changes are
the value of the output, and the decision of whether it is necessary to explore the children or 
omit these nodes. Thus, we will explain the depth-first traversal only for the intersection,
and highlight the differences for the rest of the operations.

The general idea is to process the input BRWTs column by column, recursively.
Algorithm~\ref{algorithm_llamadageneralBRWT} describes how to perform the intersection
between two BRWTs, processing every column of the root nodes calling the recursive algorithm for 
the intersection (Algorithm~\ref{algorithm_inters2_wt}). An indication of how to adapt these algorithms
to perform the difference and symmetric difference is shown later.

For the intersection, testing the value of a column in a given node, if both inputs have a $1$ 
(meaning this column is defined in both BRWTs), requires a recursive checking. 
However, if one of the inputs does not
have this column defined, the output of the intersection will be a $0$, and there is no need 
to process their children. This is done for the parts $b_l$ and $b_r$ of each node $b$.
For the intersection, the algorithm produces a column in the output if any of the 
parts ($b_l$ or $b_r$) has this column defined. Otherwise, the column is omitted in the output.

Note that, even when the access to any child could be done by using the \texttt{rank}
and \texttt{select} operations for bitmaps, we considered the use of pointers to
speed up the operations\footnote{We have included in Section~\ref{subsec:scalability} a brief note about the implications of this change.}.
% reviewer 1 Q 4 (footnote anterior) 
The $initPointersBRWT$ operation in Algorithm~\ref{algorithm_llamadageneralBRWT}
initializes these pointers to the start of each node. During the operation,
if a column is defined in one of the BRWTs but not in the other, the $Skip$ function
updates the pointers of the descendant nodes to omit this column. Otherwise, the
pointers are shifted one position to process the next column after recursively
computing the output value.

% Reviewer 1 - Comment 3 - 
Algorithm \ref{algorithm_skip} shows the $Skip$ function. Although in the worst
case it would have to process all nodes of the BRWT, this case is extremely infrequent
in practice.

%%%%%%%%%%%%%%%%%%%%%%%%%%%%%%%%%%%%%%%%%%%%%%%%%%%%%%%%%%%%
%%%%%%%%%%%% Algoritmo para Nodo Raiz %%%%%%%%%%%%%%%%%%%%%%
%%%%%%%%%%%%%%%%%%%%%%%%%%%%%%%%%%%%%%%%%%%%%%%%%%%%%%%%%%%%
\begin{algorithm}[htb!]
	\caption{\textbf{IntersectBRWT}($A,B$)}
	\scriptsize
	\algsetup{linenosize=\scriptsize}
	\begin{algorithmic}[1]
		\STATE$ pA \leftarrow initPointersBRWT(A) $
		\STATE$ pB \leftarrow initPointersBRWT(B) $
		\STATE$ bA \leftarrow 1, bB \leftarrow 1 $
		\STATE$ idNode \leftarrow 0 $
		% Reviewer 1 - Comment 1 - Se nos pasó esta línea de una versión previa en la que el algoritmo era más detallado
		%\STATE$ idFirstLeaf \leftarrow firstLeaf(numOfObjects) $
		\FOR{$i\leftarrow 0\dots numOfObjects$}
		% Reviewer 1 - Comment 2 - Falta detallar en el texto cuáles son las operaciones posibles:
		%(Intersection, Difference y SimmetricDifference.
		\STATE $Intersect(A, B, pA, pB, bA, bB, R, idNode)$ \label{RecCallWT}
		\ENDFOR
		\RETURN $R$
	\end{algorithmic}
	\label{algorithm_llamadageneralBRWT}
\end{algorithm}

%%%%%%%%%%%%%%%%%%%%%%%%%%%%%%%%%%%%%%%%%%%%%%%%%%%%%%%%%%%%
%%%%%%%%%%%%% Algoritmo Intersección %%%%%%%%%%%%%%%%%%%%%%%
%%%%%%%%%%%%%%%%%%%%%%%%%%%%%%%%%%%%%%%%%%%%%%%%%%%%%%%%%%%%
\begin{algorithm}[htb!]
	\caption{\textbf{Intersect}($A, B, pA, pB, rA, rB, R, idNode$)}
	\scriptsize
	\algsetup{linenosize=\scriptsize}
	\begin{algorithmic}[1]
		\STATE $idCh_{left} \leftarrow (idNode + 1) * 2$
		\STATE $idCh_{right} \leftarrow (idNode + 2) * 2$
		\STATE $kl \leftarrow 0, kr \leftarrow 0$ \label{wtree_inter_ceroValues}
		\STATE $bA_1 \leftarrow rA \wedge A[pA[idNode]]$
		\STATE $bA_2 \leftarrow rA \wedge A[pA[idNode + 1]]$
		\STATE $bB_1 \leftarrow rB \wedge B[pB[idNode]]$
		\STATE $bB_2 \leftarrow rB \wedge B[pB[idNode + 1]]$
		
        %\tikzmark{start01}
        \COMMENT{BEGIN code specific to Intersection}
        \IF{$!isLeaf(idNode)$} \label{line:start-specific}
			\IF{$ bA_1 \wedge bB_1 $}
				\STATE $ kl \leftarrow Intersect(A, B, pA, pB, bA_1, bB_1, R, idCh_{left}) $ \label{wtree_inter_internalNodesL}
			\ELSIF{$ bA_1 $}
				\STATE $ Skip(A, pA, idCh_{left})$ \label{wtree_inter_skipL1}
			\ELSIF{$ bB_1 $}
				\STATE$ Skip(B, pB, idCh_{left}) $ \label{wtree_inter_skipL2}
			\ENDIF
			
			\IF{$ bA_2 \wedge bB_2 $}
				\STATE $ kr \leftarrow Intersect(A, B, pA, pB, bA_2, bB_2, idCh_{right})$ \label{wtree_inter_internalNodesR}
			\ELSIF{$ bA_2 $}
				\STATE $ Skip(A, pA, idCh_{right})$ \label{wtree_inter_skipR1}
			\ELSIF{$ bB_2 $}
				\STATE$ Skip(B, pB, idCh_{right})$ \label{wtree_inter_skipR2}
			\ENDIF
    	\ELSE
    		\STATE$ kl \leftarrow bA_1 \wedge bB_1 $ \label{wtree_inter_AndOp1}
    		\STATE$ kr \leftarrow bA_2 \wedge bB_2 $ \label{wtree_inter_AndOp2}
    	\ENDIF
    	
        \label{line:end-specific} \COMMENT{END code specific to Intersection}
			
		%\tikzmark{end01}
		\IF{$ kl \vee kr \vee isRootNode(idNode) $}
			\STATE$ R[idNode] \leftarrow R[idNode] || kl$
			\STATE$ R[idNode + 1] \leftarrow R[idNode + 1] || kr$			
		\ENDIF
		\IF{$ rA \vee isRootNode(idNode) $}
			\STATE $pA[idNode] \leftarrow pA[idNode] + 1$
			\STATE $pA[idNode + 1] \leftarrow pA[idNode + 1] + 1$
		\ENDIF
		\IF{$ rB \vee isRootNode(idNode) $}
				\STATE $pB[idNode] \leftarrow pB[idNode] + 1$
				\STATE $pB[idNode+1] \leftarrow pB[idNode+1] + 1$
		\ENDIF
		\RETURN $ kl \vee kr $
	\end{algorithmic}
	\label{algorithm_inters2_wt}
	%\Textbox{start01}{end01}{Piece of code 1}
\end{algorithm}

% Reviewer 1 - Comment 3 - Falta referenciar en el texto el algoritmo skip.

%%%%%%%%%%%%%%%%%%%%%%%%%%%%%%%%%%%%%%%%%%%%%%%%%%
%%%%%%%%%%%% Algoritmo Skip %%%%%%%%%%%%%%%%%%%%%%
%%%%%%%%%%%%%%%%%%%%%%%%%%%%%%%%%%%%%%%%%%%%%%%%%%
\begin{algorithm}[htb!]
	\caption{\textbf{Skip}($X,pX, idNode$)}
	\scriptsize
	\algsetup{linenosize=\scriptsize}
	\begin{algorithmic}[1]
		\IF{$!isLeaf(idNode)$}
			\STATE $ bX_1 \leftarrow X[pX[idNode]]$
			\IF{$ bX_1 $}
			    \STATE $ idCh_{left} \leftarrow (idNode + 1) * 2$
				\STATE$ Skip(X, pX, idCh_{left})$
			\ENDIF
			\STATE $ bX_2 \leftarrow X[pX[idNode + 1]]$
			\IF{$ bX_2 $}
			    \STATE $ idCh_{right} \leftarrow (idNode + 2) * 2$
				\STATE$ Skip(X, pX, idCh_{left})$
			\ENDIF		
		\ENDIF
		\STATE $pX[idNode] \leftarrow pX[idNode] + 1$
		\STATE $pX[idNode + 1] \leftarrow pX[idNode + 1] + 1$
	\end{algorithmic}
	\label{algorithm_skip}
\end{algorithm}

As Algorithm~\ref{algorithm_inters2_wt} shows, lines \ref{line:start-specific}--\ref{line:end-specific}
are specific for the intersection. These lines must be modified to implement the difference and 
symmetric difference. The pseudocode for these changes is shown in Table~\ref{tab:codes}, but 
in summary there are basically two changes: the output bit for the column, and the management
of the column when it is defined in only one BRWT. The output bit in lines 25--26 of 
Algorithm~\ref{algorithm_inters2_wt} is computed by performing the $AND$ between the two bits 
for the intersection, while it is the $AND$ with negated second bit for the difference,
and the $EXOR$ for the symmetric difference. The recursion when the current column is defined for
both input BRWTs is the same for all the algorithms, but when only one column is defined, they differ.
We introduce a new $Copy$ function, which basically copies a column in one of the input bitmaps
to the output bitmap. 
The algorithm for the difference copies the current column of the first bitmap if it is not
defined in the second input. On the contrary, if the column is only defined in the second
bitmap, it is skipped. For the symmetric difference, if the column is defined in either bitmap,
it is copied to the output. No skipping is needed for this algorithm.

\begin{table}[t]
\scriptsize
    \begin{tabular}{|l|}
        \hline 
        \textbf{Difference} \\
        \hline
        \textit{\textbf{if}} {$!isLeaf(idNode)$} \textit{\textbf{then}}\\
		    \hspace{0.4cm} \textit{\textbf{if}}{$ bA_1 \wedge bB_1 $}\textit{\textbf{then}}\\
		        \hspace{0.8cm}$ kl \leftarrow RecDifference(A, B, pA, pB, bA_1, bB_1, R, idCh_{left}) $ \\
		    \hspace{0.4cm}\textit{\textbf{else if}} {$ bA_1 $} \textit{\textbf{then}}\\
		        \hspace{0.8cm}$ Copy(A, pA, R, idCh_{left})$\\
		        \hspace{0.8cm}$kl \leftarrow 1$\\
		    \hspace{0.4cm}\textit{\textbf{else if}} {$ bB_1 $} \textit{\textbf{then}}\\
    		    \hspace{0.8cm}$ Skip(B, pB, idCh_{left}) $ \\
            \hspace{0.4cm}\textit{\textbf{end if}}
		
    		\hspace{0.4cm}\textit{\textbf{if}} {$ bA_2 \wedge bB_2 $} \textit{\textbf{then}}\\
    		    \hspace{0.8cm}$ kr \leftarrow RecDifference(A, B, pA, pB, bA_2, bB_2, idCh_{right})$\\ 
    		\hspace{0.4cm}\textit{\textbf{else if}} {$ bA_2 $} \textit{\textbf{then}}\\
    		    \hspace{0.8cm}$ Copy(A, pA, R, idCh_{right})$ \\
    		\hspace{0.4cm}\textit{\textbf{else if}} {$ bB_2 $} \textit{\textbf{then}}\\
    		    \hspace{0.8cm}$ Skip(B, pB, idCh_{right})$ \\
    		\hspace{0.4cm}\textit{\textbf{end if}}\\
    	\textit{\textbf{else}}\\
		    \hspace{0.4cm}$ kl \leftarrow bA_1 \wedge \sim bB_1 $\\
		    \hspace{0.4cm}$ kr \leftarrow bA_2 \wedge \sim bB_2 $\\
		\textit{\textbf{end if}}
        \\ \\ \hline \hline
        \textbf{Symmetric difference} \\
        \hline
        \textit{\textbf{if}} {$!isLeaf(idNode)$} \textit{\textbf{then}}\\
        	\hspace{0.4cm}\textit{\textbf{if}} {$ bA_1 \wedge bB_1 $} \textit{\textbf{then}}\\
        		\hspace{0.8cm}$ kl \leftarrow RecSymmDiff(A, B, pA, pB, bA_1, bB_1, R, idCh_{left}) $ \\
        	\hspace{0.4cm}\textit{\textbf{else if}} {$ bA_1 $} \textit{\textbf{then}}\\
        		\hspace{0.8cm}$ Copy(A, pA, R, idCh_{left})$\\
        	\hspace{0.4cm}\textit{\textbf{else if}} {$ bB_1 $} \textit{\textbf{then}}\\
        		\hspace{0.8cm}$ Copy(B, pB, R, idCh_{left}) $\\
        	\hspace{0.4cm}\textit{\textbf{else}}\\
        		\hspace{0.8cm}$ kl \leftarrow 0$ \\
        	\hspace{0.4cm}\textit{\textbf{end if}}\\
        	
        	\hspace{0.4cm}\textit{\textbf{if}} {$ bA_2 \wedge bB_2 $} \textit{\textbf{then}}\\
        		\hspace{0.8cm}$ kr \leftarrow RecSymmDiff(A, B, pA, pB, bA_2, bB_2, idCh_{right})$ \\	
        		\hspace{0.4cm}\textit{\textbf{else if}} {$ bA_2 $} \textit{\textbf{then}}\\
        		\hspace{0.8cm}$ Copy(A, pA, R, idCh_{right})$\\
        	\hspace{0.4cm}\textit{\textbf{else if}} {$ bB_2 $} \textit{\textbf{then}}\\
        		\hspace{0.8cm}$ Copy(B, pB, R, idCh_{right})$\\
        	\hspace{0.4cm}\textit{\textbf{else}}\\
        		\hspace{0.8cm}$ kr \leftarrow 0$ \\
        	\hspace{0.4cm}\textit{\textbf{end if}}\\
        \textit{\textbf{else}}\\
        	\hspace{0.4cm}$ kl \leftarrow (\sim bA_1 \wedge bB_1) \vee (bA_1 \wedge \sim bB_1) $ \\
        	\hspace{0.4cm}$ kr \leftarrow (\sim bA_2 \wedge bB_2) \vee (bA_2 \wedge \sim bB_2) $ \\
        \textit{\textbf{end if}}
        \\ \\ \hline
    \end{tabular}
    \caption{Code snippets for set operations.}
    \label{tab:codes}
\end{table}
\normalsize

\section{Empirical Evaluation}\label{sec:experiments}
In this section we describe the experiments we have conducted to compare the performance of the three compact data structures ($k^2$-trees,
$k^2$-tree1s, and BRWT) and that of the compressed adjacency lists used to represent
binary relations.

We first describe the datasets used in our experiments, of which one of them is real and
three of them are synthetic. Then, we include some implementation details of our algorithms, and
describe the experimental hardware and software framework we have used. Finally, we present our results.

\subsection{Datasets}
We ran our experiments over a real dataset (\texttt{snaps-uk})
and three synthetically generated distributions that use well-known 
random models, such as Erd{\H o}s and R{\'e}nyi  \cite{randomgraphs:1959}, 
small-world (using Newman Watts-Strogatz distribution \cite{smallworld:2001}),
and Barabasi-Albert distribution \cite{barabasi:2002}. We shall refer to these
dataset distributions as \texttt{random}, \texttt{smallworld} and \texttt{barabasi},
respectively. Each dataset is formed by 12 files, so the metrics obtained on the
sizes and timings consider the average for these files.

The \texttt{snaps-uk} dataset was taken from a series of twelve monthly snapshots 
of a Web graph from the \texttt{.uk} domain, collected by the Laboratory for Web 
Algorithmics\footnote{\url{http://law.di.unimi.it}} \cite{BoVWFI,BRSLLP,BSVLTAG}. 
We have cut down these graphs to use 1 million nodes ($n = 1,000,000$). More
information on this dataset is also available in \cite{QUIJADAFUENTES201976}.

The three synthetic datasets were generated using the 
NetworkX\footnote{\url{https://networkx.github.io/}} Python library. For the 
random distribution, only the number of nodes $n$ and number of edges $m$ were required
as arguments. For the Barabasi-Albert and small-world distributions, a third parameter
$k$ is needed, to specify the $k$-nearest neighbors to connect to a given node. This 
parameter determines the number of edges that is generated. If the number of generated
edges is greater than the specified number $m$, the remaining edges are removed.

The output is a binary file containing the plain adjacency list, which in turn is used 
to build the compact data structure representations (for $k^2$-trees, $k^2$-tree1s, 
and BRWT) as well as the compressed adjacency lists (QMX and Rice-runs).

We have also generated 12 files for each distribution. All of them have $n = 1,000,000$ nodes.
In order to have the same density, the number of generated
edges for each file was the same as the corresponding file for the \texttt{snaps-uk} dataset.
That resulted in an average number of edges of $m = 2,240,877$ per distribution, which gives
a density $m/n^2 = 0.000224\%$.

\begin{figure*}[t]
    \centering
    \begin{subfigure}[Barabasi]{0.45\textwidth}{\includegraphics[width=.98\textwidth]{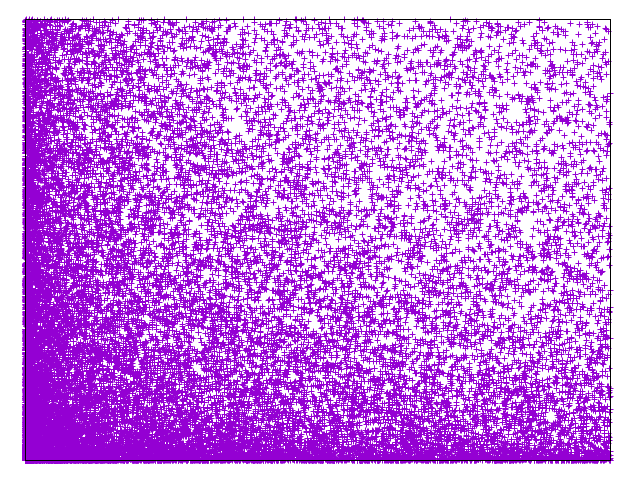}}
    \caption{\texttt{Barabasi}}
    \end{subfigure}
    \begin{subfigure}[Random]{0.45\textwidth}{\includegraphics[width=.98\textwidth]{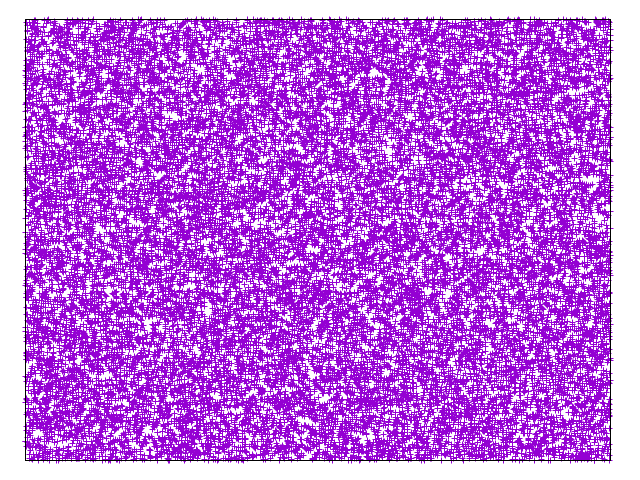}}
        \caption{\texttt{random}}
    \end{subfigure}
    \begin{subfigure}[Smallworld]{0.45\textwidth}{\includegraphics[width=.98\textwidth]{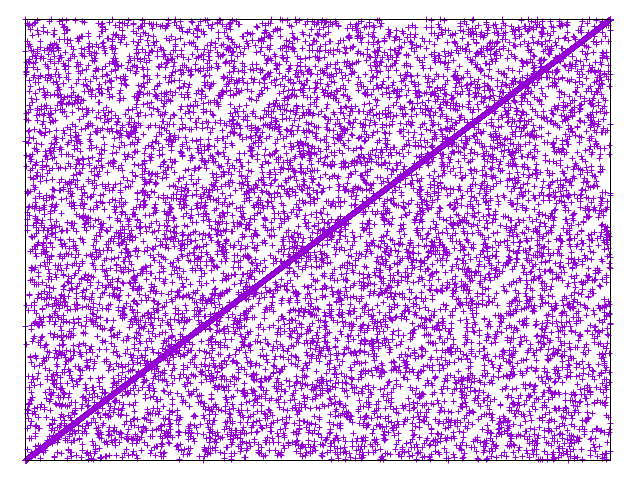}}
        \caption{\texttt{smallworld}}
    \end{subfigure}
    \begin{subfigure}[Snaps-uk]{0.45\textwidth}{\includegraphics[width=.98\textwidth]{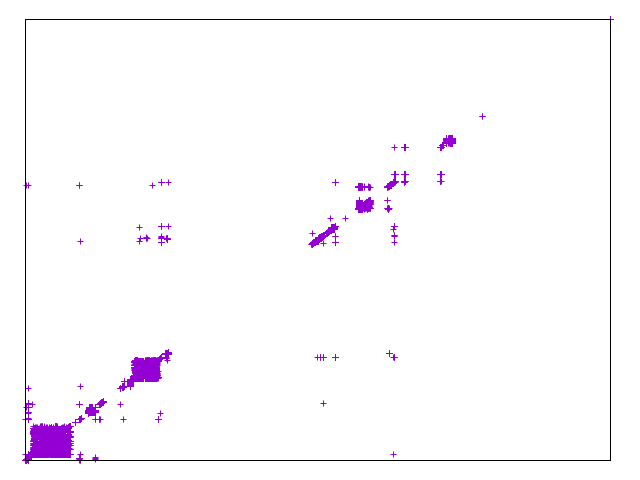}}
    \caption{\texttt{snaps-uk}}
    \end{subfigure}
    \caption{A sample of all dataset distributions.}
    \label{fig:distributions}
\end{figure*}

Given that the distribution of 1s has a high impact on the size of the compressed structures,
as well as in the performance of data structures, a sample of all datasets is shown in
Figure~\ref{fig:distributions}. As we can see, the synthetically generated datasets are much
less clustered than \texttt{snaps-uk}, the real dataset.

\subsection{Experimental framework}
The comparison we performed considered, for all the data structures, the following operations:
\begin{itemize}
    \item Neighborhood queries: isRelated, %(find whether there is a 1 or a 0 in $(x,y)$ coordinates
    %of the matrix, 
    successors, predecessors, and range neighborhood.
    \item Set operations: union, intersection, difference, and symmetric difference.
\end{itemize}

The work described here required the coding of all of the algorithms (both neighborhood and set operations) for BRWT, as well as the neighborhood operations for $k^2$-tree1s and compressed adjacency
lists using QMX and Rice-runs. For the remaining algorithms, we use the source code by the authors of the works described in \cite{BLN14} and \cite{BAB+13}.

The implementation language for all algorithms is C, compiled with gcc version 6.3.0. 
The experiments were run on an isolated Intel\textsuperscript{\textregistered} 
Xeon\textsuperscript{\textregistered} ES2470@2.30GHz processor with 20 MB of cache, 
and 64 GB of RAM. It runs Debian 10.1 (buster) with kernel 4.19.0 (64 bits).

\subsection{Results}

We present in this section the results of our empirical evaluation. First, we study 
independently the storage needs (the actual size) of the data 
structures, and their performance for both neighborhood queries and set operations. 
Then, both sizes and times are considered together in order to present some trade-offs
that would apply when choosing a specific data structure. Finally, we analyze the 
data structures in terms of scalability.

In the neighborhood operations, besides the isRelated operation, we have included a similar
one: isRelated-True. In fact, it is the same operation, but the result of the query
is known to be $true$ (which corresponds to having a 1 in the matrix). Thus, isRelated-True acts as a worst-case scenario
for the isRelated operation in most cases. For example, for $k^2$-trees, we know that
this operation must navigate the tree until its leaves (the result is 1, so it cannot be
discarded by a 0 in a previous level of the tree).

Also note that, for the naming of the data structures in the tables and graphics of this
section, we have chosen shorter names: \texttt{kt} for $k^2$-tree, \texttt{ktone} for
$k^2$-tree1, \texttt{brwt} for BWRT, \texttt{qmx} for the QMX encoder, and
\texttt{rice} for the Rice-runs encoder.

\subsubsection{Storage}
The average size taken up by each structure for all distribution is shown in Table~\ref{tab:sizes:bytes}. 
As a reference, the size of the full uncompressed adjacency lists is also included in the table. Our standard
implementation uses 32-bit integers, so it is used as the base number. However, in order to represent a relation
for 1 million nodes, only 20 bits suffice. Thus, we also show the size theoretically needed to represent 
this relation. Table~\ref{tab:sizes:ratio} shows the same information as a ratio, considering 
the full adjacency lists as the base for comparison (value $1.0$), so the deviations can be better seen.

\begin{table*}[t]
    \centering
\begin{tabular}{|l|r|r|r|r|}
\hline
 & \texttt{barabasi}  & \texttt{random} & \texttt{smallworld} & \texttt{snaps-uk}
\\\hline
  Full adj. list (32 bits)  & 12,963,523 & 12,963,523 & 12,963,523 & 12,963,523 \\\hline
  Full adj. list (20 bits)  & 8,102,201 & 8,102,201 & 8,102,201 & 8,102,201 \\\hline
  \texttt{brwt} & 10,367,083 & 10,894,448 & 7,413,171 & 2,371,233 \\\hline
  \texttt{kt} & 9,737,915 & 10,724,430 & 4,770,940 & 1,419,065 \\\hline
  \texttt{ktone} & 15,613,223 & 17,304,302 & 7,344,613 & 1,771,421 \\\hline
  \texttt{qmx} & 16,567,964 & 22,101,395 & 19,599,142 & 7,569,547 \\\hline
  \texttt{rice} & 13,808,180 & 17,651,437 & 14,440,775 & 6,672,224 \\
\hline\end{tabular}
    \caption{Average size (in bytes) for the different datasets.}
    \label{tab:sizes:bytes}
\end{table*}

\begin{table*}[t]
    \centering
    \begin{tabular}{|l|r|r|r|r|}
\hline
 & \texttt{barabasi}  & \texttt{random} & \texttt{smallworld} & \texttt{snaps-uk}
\\\hline
  Full adj. list (32 bits) & 1.00 & 1.00 & 1.00 & 1.00 \\\hline
  Full adj. list (20 bits) & 0.63 & 0.63 & 0.63 & 0.63 \\\hline
  \texttt{brwt} & 0.80 & 0.84 & 0.57 & 0.18 \\\hline
  \texttt{kt} & 0.75 & 0.83 & 0.37 & 0.11 \\\hline
 \texttt{ktone} & 1.20 & 1.33 & 0.57 & 0.14 \\\hline
  \texttt{qmx} & 1.28 & 1.70 & 1.51 & 0.58 \\\hline
 \texttt{rice} & 1.07 & 1.36 & 1.11 & 0.51 \\
\hline\end{tabular}
    \caption{Average size, shown as a ratio.}
    \label{tab:sizes:ratio}
\end{table*}

Considering the datasets, we can see that the distribution that allows for the best
compression ratios (actually, the only one that gets compressed by all
structures) is \texttt{snaps-uk}. This is reasonable, because this
distribution is clustered, unlike the three synthetic ones, which are
based on random models. More concretely, the best compression ratios are obtained by the $k^2$-tree variants, which benefit from distributions of small number of ones that are clustered. 
As for the QMX and Rice-runs, their behavior is worse, because
they are based on run-length compression, and having a smaller runs of 1s
produces worse compression ratios.

Considering the compressed data structures, we can see that standard $k^2$-trees obtain better results than the plain adjacency list using 20 bits for those clustered distributions (\texttt{smallworld} and \texttt{snaps-uk}), but not for those that follow a more random model (\texttt{barabasi} and \texttt{random}). In any case, compressed data structures obtain generally better results than compressed adjacency lists (except in the case of \texttt{barabasi}, which obtains better results for Rice-runs than for $k^2$-tree1).
Compressed adjacency lists require, for all synthetic data distributions, larger spaces than the original plain representation.
%compress the matrix (their ratio in Table~\ref{tab:sizes:ratio} 
%is always less than one), while the rest of the structures can obtain a file
%much larger than the original. 
For example, using QMX to compress the \texttt{random}
distribution actually obtains a file 70\% bigger than the full adjacency lists using 32 bits.

\subsubsection{Timings}\label{subsec:timings}
The timings shown in this section consider only the time devoted to the
operations themselves, without taking into account the I/O time of reading
the structures (for neighborhood and set operations) or writing the result
(only for set operations). In the case of the neighborhood operations, the time
shown corresponds to the execution of $1,000$ queries of the same type over
the same structure.

Let us first discuss the neighborhood operations. Their timings, in milliseconds,
are shown in Table~\ref{tab:times:search:ms}. For a better comparison,
Table~\ref{tab:times:search:ratio} shows the same information as ratios.
The value $1.00$ corresponds to the shortest time for the operation on a 
distribution.

\begin{table*}[t]
    \centering
\begin{tabular}{|l|l|r|r|r|r|}
\hline
Query & & \texttt{barabasi}  & \texttt{random} & \texttt{smallworld} & \texttt{snaps-uk} 
\\\hline
isRelated &   \texttt{brwt} & 0.312 & 0.418 & 0.261 & 0.075 \\
    &      \texttt{kt} & 0.975 & 1.240 & 1.057 & 0.313 \\
    &      \texttt{ktone} & 0.867 & 1.192 & 1.122 & 0.453 \\
    &      \texttt{qmx} & 0.565 & 0.941 & 0.602 & 0.162 \\
    &      \texttt{rice} & 0.460 & 0.586 & 0.456 & 0.135 \\\hline
isRelated-True & \texttt{brwt} & 4.539 & 4.563 & 4.607 & 6.103 \\
    &      \texttt{kt} & 2.768 & 2.601 & 2.620 & 3.318 \\
    &      \texttt{ktone} & 2.899 & 3.501 & 2.506 & 3.260 \\
    &      \texttt{qmx} & 1.461 & 1.121 & 1.103 & 3.690 \\
    &      \texttt{rice} & 1.335 & 0.655 & 0.645 & 3.023 \\\hline
predecessors & \texttt{brwt} & 19.814 & 20.709 & 16.269 & 14.018 \\
    &      \texttt{kt} & 270.140 & 356.439 & 190.632 & 11.515 \\
    &      \texttt{ktone} & 444.631 & 653.985 & 304.679 & 14.634 \\
    &      \texttt{qmx} & 337111.835 & 468610.372 & 358252.193 & 101397.174 \\
    &      \texttt{rice} & 206217.818 & 270400.268 & 200732.389 & 78405.505 \\\hline
successors & \texttt{brwt} & 16.258 & 18.792 & 16.278 & 16.322 \\
    &      \texttt{kt} & 263.515 & 334.947 & 189.781 & 10.459 \\
    &      \texttt{ktone} & 437.586 & 635.259 & 298.441 & 13.610 \\
    &      \texttt{qmx} & 0.660 & 0.841 & 0.755 & 0.269 \\
    &      \texttt{rice} & 0.461 & 0.641 & 0.482 & 0.195 \\\hline
rangeNeighborhood & \texttt{brwt} & 35.709 & 40.745 & 32.419 & 21.644 \\
    &      \texttt{kt} & 1.767 & 2.461 & 1.371 & 0.316 \\
    &      \texttt{ktone} & 4.308 & 6.175 & 1.860 & 0.373 \\
    &      \texttt{qmx} & 80.690 & 110.513 & 124.586 & 53.651 \\
    &      \texttt{rice} & 75.017 & 88.380 & 76.048 & 48.724 \\
\hline\end{tabular}
    \caption{Timings for neighborhood queries (in ms).}
    \label{tab:times:search:ms}
\end{table*}

\begin{table*}[t]
    \centering
\begin{tabular}{|l|l|r|r|r|r|}
\hline
Query & & \texttt{barabasi}  & \texttt{random} & \texttt{smallworld} & \texttt{snaps-uk} 
\\\hline
isRelated & \texttt{brwt} & 1.00 & 1.00 & 1.00 & 1.00 \\
    &      \texttt{kt} & 3.13 & 2.97 & 4.05 & 4.17 \\
    &      \texttt{ktone} & 2.78 & 2.85 & 4.30 & 6.04 \\
    &      \texttt{qmx} & 1.81 & 2.25 & 2.31 & 2.16 \\
    &      \texttt{rice} & 1.47 & 1.40 & 1.75 & 1.80 \\\hline
isRelated-True & \texttt{brwt} & 3.40 & 6.97 & 7.14 & 2.02 \\
    &      \texttt{kt} & 2.07 & 3.97 & 4.06 & 1.10 \\
    &      \texttt{ktone} & 2.17 & 5.35 & 3.89 & 1.08 \\
    &      \texttt{qmx} & 1.09 & 1.71 & 1.71 & 1.22 \\
    &      \texttt{rice} & 1.00 & 1.00 & 1.00 & 1.00 \\\hline
predecessors & \texttt{brwt} & 1.00 & 1.00 & 1.00 & 1.22 \\
    &      \texttt{kt} & 13.63 & 17.21 & 11.72 & 1.00 \\
    &      \texttt{ktone} & 22.44 & 31.58 & 18.73 & 1.27 \\
    &      \texttt{qmx} & 17013.82 & 22628.34 & 22020.54 & 8805.66 \\
    &      \texttt{rice} & 10407.68 & 13057.14 & 12338.34 & 6808.99 \\\hline
successors & \texttt{brwt} & 35.27 & 29.32 & 33.77 & 83.70 \\
    &      \texttt{kt} & 571.62 & 522.54 & 393.74 & 53.64 \\
    &      \texttt{ktone} & 949.21 & 991.04 & 619.17 & 69.79 \\
    &      \texttt{qmx} & 1.43 & 1.31 & 1.57 & 1.38 \\
    &      \texttt{rice} & 1.00 & 1.00 & 1.00 & 1.00 \\\hline
rangeNeighborhood & \texttt{brwt} & 20.21 & 16.56 & 23.65 & 68.49 \\
    &      \texttt{kt} & 1.00 & 1.00 & 1.00 & 1.00 \\
    &      \texttt{ktone} & 2.44 & 2.51 & 1.36 & 1.18 \\
    &      \texttt{qmx} & 45.66 & 44.91 & 90.87 & 169.78 \\
    &      \texttt{rice} & 42.45 & 35.91 & 55.47 & 154.19 \\
\hline\end{tabular}
    \caption{Timings for neighborhood queries (ratio).}
    \label{tab:times:search:ratio}
\end{table*}

The information that stands out most in these tables corresponds to the 
\emph{predecessors} operation, where the fastest structure is the
BRWT, and the QMX and Rice-runs encoders are much slower (up to $22,628$ 
times slower in the case of the \texttt{random} dataset). This is reasonable because encoders
compress the adjacency lists row by row, and finding the predecessors requires 
the decompression of all of the encoded lists. 

For \emph{successors}, however, Rice-runs is the fastest, closely
followed by QMX, while $k^2$-trees and $k^2$-tree1s are slower (almost $1,000$ times).
The reason is that, in this case, the encoders have to decompress only one list
(at most; if there are no 1s in the row, the answer is immediate). 

Considering together predecessors and successors, we can see that the difference
between the best and the worst is much larger in the predecessors because, as
we mentioned, QMX and Rice-runs must decode all of the lists. However, for successors,
the compact data structures have to explore only part of the binary relation, not all of
it. This is because all of them are actually self-indices, so they allow for a fast access
to a portion of the matrix. 
For the same reason, if we consider the \emph{rangeNeighborhood} queries, we can see that the compact
data structures perform better than the encoders, and, in this case, $k^2$-tree is the
fastest structure.

For the \emph{isRelated} and \emph{isRelated-True} queries, all structures offer a more
homogeneous behavior. Anyway, BRWT is the fastest for \emph{isRelated}, while it is Rice-runs 
for \emph{isRelated-True}. This is due to the fact that \emph{isRelated} accesses a random
cell in the matrix, and with a low density BRWT is able to answer {\it false} without
reaching the leaf nodes, while in the case of \emph{isRelated-True} query, BRWT must always reach a leaf node. 
The same happens for $k^2$-trees and $k^2$-tree1s.

\begin{table*}[t]
    \centering
\begin{tabular}{|l|l|r|r|r|r|}
\hline
Operation & & \texttt{barabasi}  & \texttt{random} & \texttt{smallworld} & \texttt{snaps-uk} 
\\\hline
Difference & \texttt{brwt} & 8228.666 & 9336.838 & 3706.757 & 745.321 \\
    &      \texttt{kt} & 4244.713 & 5001.098 & 1942.896 & 349.290 \\
    &      \texttt{ktone} & 5468.160 & 6072.274 & 2188.450 & 265.957 \\
    &      \texttt{qmx} & 715.745 & 890.225 & 694.532 & 310.097 \\
    &      \texttt{rice} & 480.035 & 563.134 & 377.019 & 148.794 \\\hline
Intersection & \texttt{brwt} & 3887.119 & 4313.796 & 2304.365 & 610.479 \\
    &      \texttt{kt} & 2543.189 & 3095.668 & 1233.486 & 199.862 \\
    &      \texttt{ktone} & 4958.742 & 5491.372 & 2040.778 & 224.208 \\
    &      \texttt{qmx} & 562.268 & 682.922 & 680.239 & 321.681 \\
    &      \texttt{rice} & 284.395 & 335.176 & 367.698 & 153.825 \\\hline
Symmetric & \texttt{brwt} & 11817.572 & 17987.314 & 5173.330 & 767.283 \\
Difference &  \texttt{kt} & 5778.197 & 6716.642 & 2676.058 & 527.112 \\
    &      \texttt{ktone} & 5317.761 & 6253.623 & 2385.553 & 379.348 \\
    &      \texttt{qmx} & 819.924 & 952.049 & 724.674 & 346.717 \\
    &      \texttt{rice} & 561.967 & 661.309 & 445.186 & 126.816 \\\hline
Union & \texttt{brwt} & 7371.075 & 7960.348 & 4201.526 & 1218.818 \\
    &      \texttt{kt} & 4117.581 & 4517.057 & 1927.165 & 571.612 \\
    &      \texttt{ktone} & 5552.810 & 6351.123 & 2392.387 & 444.056 \\
    &      \texttt{qmx} & 841.915 & 999.932 & 780.056 & 410.269 \\
    &      \texttt{rice} & 624.011 & 680.933 & 524.918 & 271.250 \\
\hline\end{tabular}
    \caption{Timings for set operations (in ms).}
    \label{tab:times:set:ms}
\end{table*}

\begin{table*}[t]
    \centering
\begin{tabular}{|l|l|r|r|r|r|}
\hline
Operation & & \texttt{barabasi}  & \texttt{random} & \texttt{smallworld} & \texttt{snaps-uk} 
\\\hline
Difference & \texttt{brwt} & 17.14 & 16.58 & 9.83 & 5.01 \\
    &      \texttt{kt} & 8.84 & 8.88 & 5.15 & 2.35 \\
    &      \texttt{ktone} & 11.39 & 10.78 & 5.80 & 1.79 \\
    &      \texttt{qmx} & 1.49 & 1.58 & 1.84 & 2.08 \\
    &      \texttt{rice} & 1.00 & 1.00 & 1.00 & 1.00 \\\hline
Intersection & \texttt{brwt} & 13.67 & 12.87 & 6.27 & 3.97 \\
    &      \texttt{kt} & 8.94 & 9.24 & 3.35 & 1.30 \\
    &      \texttt{ktone} & 17.44 & 16.38 & 5.55 & 1.46 \\
    &      \texttt{qmx} & 1.98 & 2.04 & 1.85 & 2.09 \\
    &      \texttt{rice} & 1.00 & 1.00 & 1.00 & 1.00 \\\hline
Symmetric & \texttt{brwt} & 21.03 & 27.20 & 11.62 & 6.05 \\
Difference    &      \texttt{kt} & 10.28 & 10.16 & 6.01 & 4.16 \\
    &      \texttt{ktone} & 9.46 & 9.46 & 5.36 & 2.99 \\
    &      \texttt{qmx} & 1.46 & 1.44 & 1.63 & 2.73 \\
    &      \texttt{rice} & 1.00 & 1.00 & 1.00 & 1.00 \\\hline
Union & \texttt{brwt} & 11.81 & 11.69 & 8.00 & 4.49 \\
    &      \texttt{kt} & 6.60 & 6.63 & 3.67 & 2.11 \\
    &      \texttt{ktone} & 8.90 & 9.33 & 4.56 & 1.64 \\
    &      \texttt{qmx} & 1.35 & 1.47 & 1.49 & 1.51 \\
    &      \texttt{rice} & 1.00 & 1.00 & 1.00 & 1.00 \\
\hline\end{tabular}
    \caption{Timings for set operations (ratio).}
    \label{tab:times:set:ratio}
\end{table*}

For the set operations, Table~\ref{tab:times:set:ms}
shows the actual times of our experiments, and the same information 
as ratios is shown in Table~\ref{tab:times:set:ratio}.
Even when the difference between the best and worst
data structure is not as large as for neighborhood operations, it is clear that the encoders
(Rice-runs, closely followed by QMX) are the best option. 
The BRWT is almost always the slowest for these kinds of operations.

The reason behind that behavior is that set operations must access in general
all the elements of the binary relation. The encoders just decompress row by
row, build the result and encode it as a new output row. However, the three
compact data structures, being self-indices, have an overhead that (as in
general for any type of index) worsens the performance when the full dataset has to
be accessed.

\subsubsection{Storage size versus time}
\begin{figure*}[t]
    \centering
    \begin{subfigure}[Neighborhood operations]{0.45\textwidth}{\includegraphics[width=.98\textwidth]{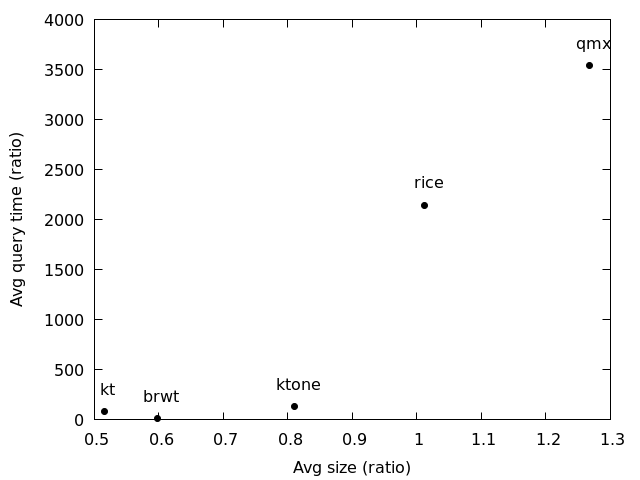} }
    \caption{Neighborhood operations}
    \end{subfigure}
    \begin{subfigure}[Set operations]{0.45\textwidth}{\includegraphics[width=.98\textwidth]{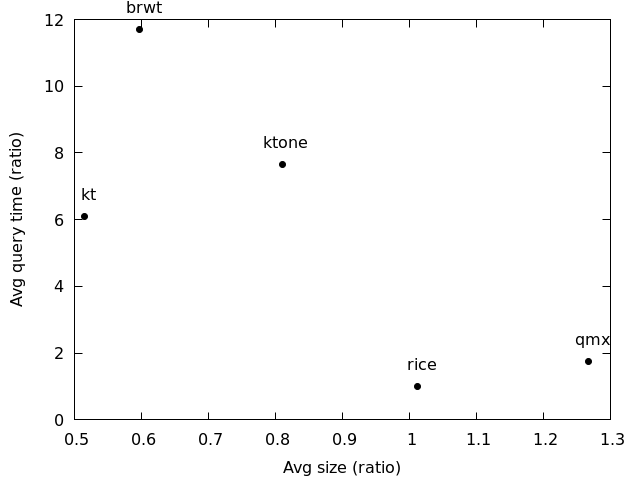}}
    \caption{Set operations}
\end{subfigure}
    \caption{Average time versus average size for all distributions.}
    \label{fig:time-size:global}
\end{figure*}

\begin{figure*}[t]
    \centering
    \begin{subfigure}[Neighborhood operations]{0.45\textwidth}{\includegraphics[width=.98\textwidth]{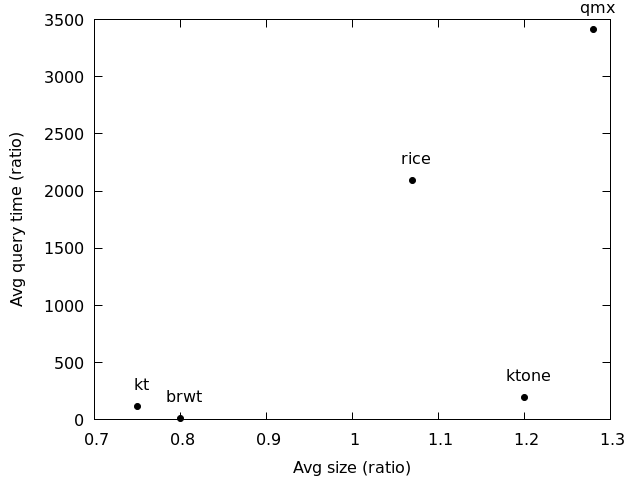}} 
    \caption{Neighborhood operations}
    \end{subfigure}
    \begin{subfigure}[Set operations]{0.45\textwidth}{\includegraphics[width=.98\textwidth]{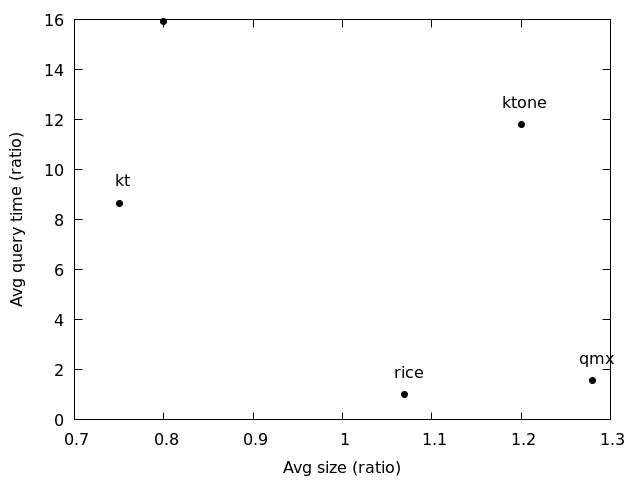}}
    \caption{Set operations}
    \end{subfigure}
    \caption{Average time versus average size for \texttt{barabasi}.}
    \label{fig:time-size:barabasi}
\end{figure*}

\begin{figure*}[t]
    \centering
    \begin{subfigure}[Neighborhood operations]{0.45\textwidth}{\includegraphics[width=.98\textwidth]{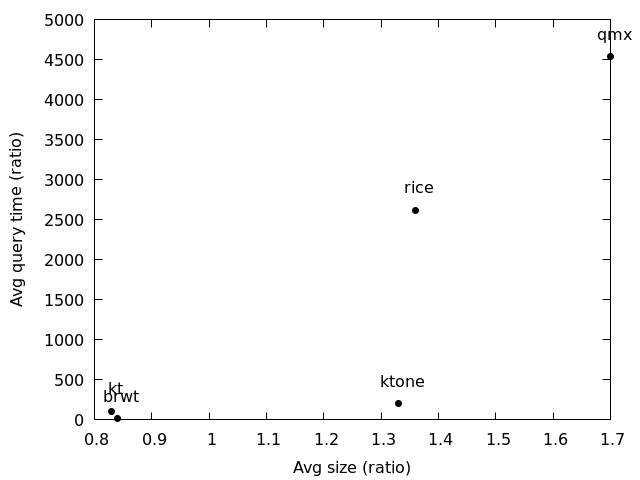}} 
    \caption{Neighborhood operations}
    \end{subfigure}
    \begin{subfigure}[Set operations]{0.45\textwidth}{\includegraphics[width=.98\textwidth]{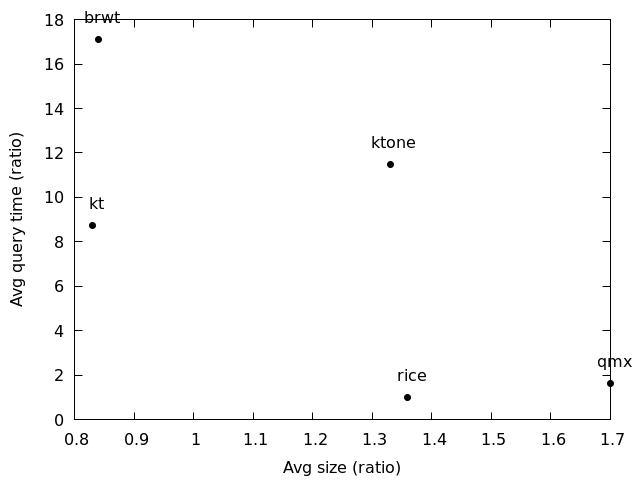} }
    \caption{Set operations}
    \end{subfigure}
    \caption{Average time versus average size for \texttt{random}.}
    \label{fig:time-size:random}
\end{figure*}

\begin{figure*}[t]
    \centering
    \begin{subfigure}[Neighborhood operations]{0.45\textwidth}{\includegraphics[width=.98\textwidth]{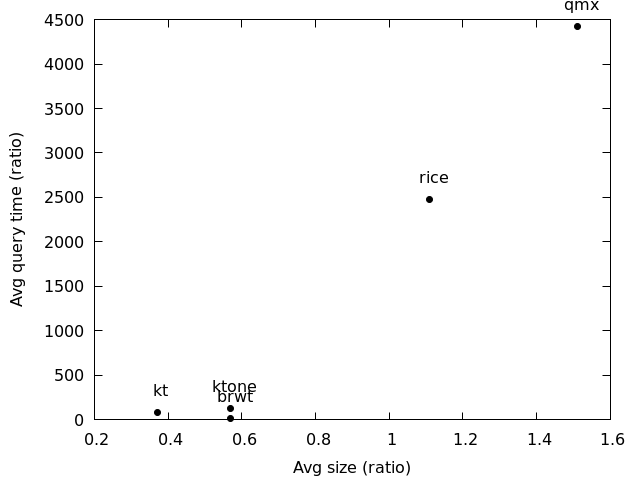}} 
    \caption{Neighborhood operations}
    \end{subfigure}
    \begin{subfigure}[Set operations]{0.45\textwidth}{\includegraphics[width=.98\textwidth]{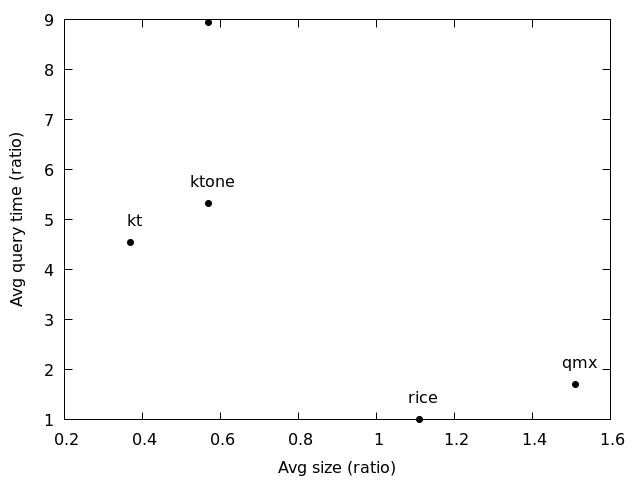} }
    \caption{Set operations}
    \end{subfigure}
    \caption{Average time versus average size for \texttt{smallworld}.}
    \label{fig:time-size:smallworld}
\end{figure*}

\begin{figure*}[t]
    \centering
    \begin{subfigure}[Neighborhood operations]{0.45\textwidth}{\includegraphics[width=.98\textwidth]{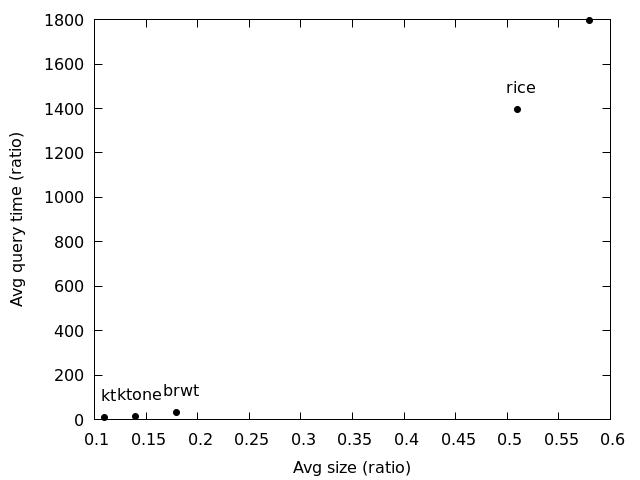}}
    \caption{Neighborhood operations}
    \end{subfigure}
    \begin{subfigure}[Set operations]{0.45\textwidth}{\includegraphics[width=.98\textwidth]{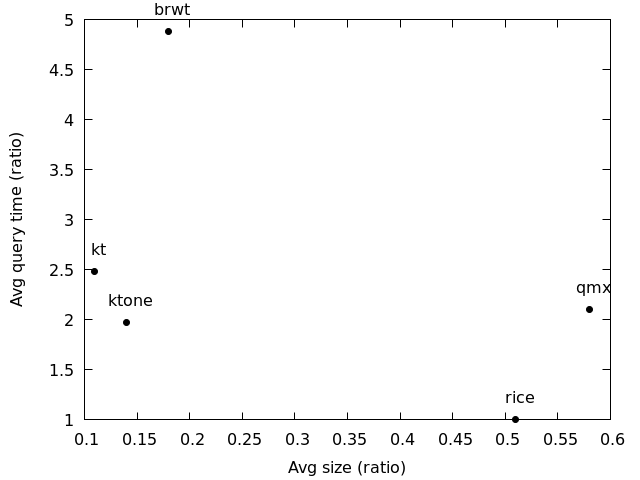}}
    \caption{Set operations}
    \end{subfigure}
    \caption{Average time versus average size for \texttt{snaps-uk}.}
    \label{fig:time-size:snaps}
\end{figure*}

Let us consider now the trade-off between storage size and performance for all compared
data structures. Figures \ref{fig:time-size:global}--\ref{fig:time-size:snaps} analyze their
behavior. All figures contain two graphics: $(a)$ for the neighborhood operations, and $(b)$ for the set 
operations.

For the neighborhood operations, it is clear that the compact data structures, especially the
standard $k^2$-tree, is the best option, in terms of both size and performance, 
while the encoders use more space and perform worse. Figure~\ref{fig:time-size:global}
shows this behavior considering an average of all the datasets together. 
If we take into account the data distribution, we can see that the previous conclusion is 
generally true, except for the $k^2$-tree1. This data structure is highly dependant
on the degree of clustering (remember that it compresses areas of 1s in the matrix),
and thus its size grows for non-clustered datasets (like \texttt{barabasi}, 
and \texttt{random}, as seen in Figures \ref{fig:time-size:barabasi} and \ref{fig:time-size:random} 
respectively), and it behaves better for the more clustered (\texttt{smallworld} and
and especially \texttt{snaps-uk}, as seen in Figures \ref{fig:time-size:smallworld} and 
\ref{fig:time-size:snaps} respectively).

For the set operations, none of the data structures
outperforms the rest in terms of size and performance. On the contrary, we can see in 
Figure~\ref{fig:time-size:global} that we have a trade-off, because the compact data
structures use less space, but they are slower than the encoders. The encoders are
faster, but they need more space. Thus, the general advice would be the following:
if we are primarily interested in speed, and there is enough available RAM to fit 
the data structures using the encoders, then use the encoders. If the datasets would
not fit into RAM, use the compact data structures. 

Parts $(b)$ of Figures \ref{fig:time-size:barabasi}--\ref{fig:time-size:snaps} show 
this behavior for each data distribution. 
We can see that the BRWT is a poor choice
for the set operations in any distribution in terms of speed (not in terms of space).
Again, the $k^2$-tree1 shows a behavior highly dependent on the clusterization, 
being one of the fastest data structures for the \texttt{snaps-uk} dataset 
and one of the slowest technique for \texttt{barabasi}.

\subsection{A note on scalability}\label{subsec:scalability}
The previous section analyzed the performance of the data structures over relations
having 1 million nodes. However, we are also interested in the behavior of the data
structures when the size of the relation grows. 

We shall describe in this section the behavior of the data structures when the number
of nodes varies, growing up to $10,000,000$ nodes. We have chosen the \texttt{smallworld}
data distribution for these experiments, because it is the less biased distribution:
it is not as clustered as the real dataset (\texttt{snaps-uk}), which would benefit
the compact data structures, and it is not as evenly distributed as \texttt{random}
or \texttt{barabasi}, which would benefit the QMX and Rice-runs encoders. The dataset
was generated also using the NetworkX Python library, considering a value of
2 for the $k$ parameter (indicating the $k$ nearest neighbors that would be linked 
in the graph) in all cases.

\begin{figure*}[t]
    \centering
    \begin{subfigure}[isRelated]{0.45\textwidth}{\includegraphics[width=.98\textwidth]{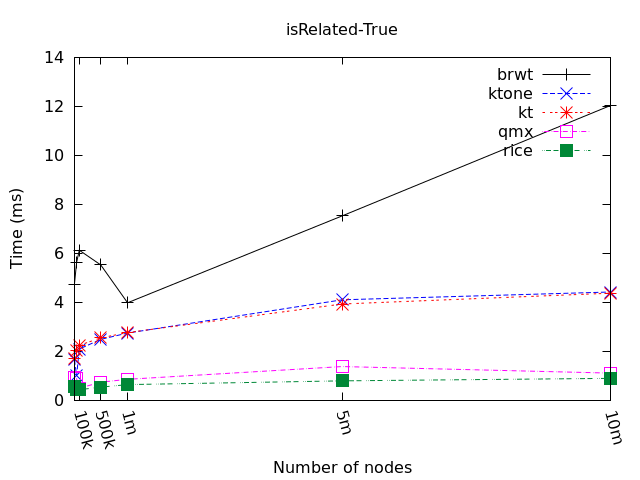}}
    \caption{isRelated}
    \end{subfigure}
    \begin{subfigure}[successors]{0.45\textwidth}{\includegraphics[width=.98\textwidth]{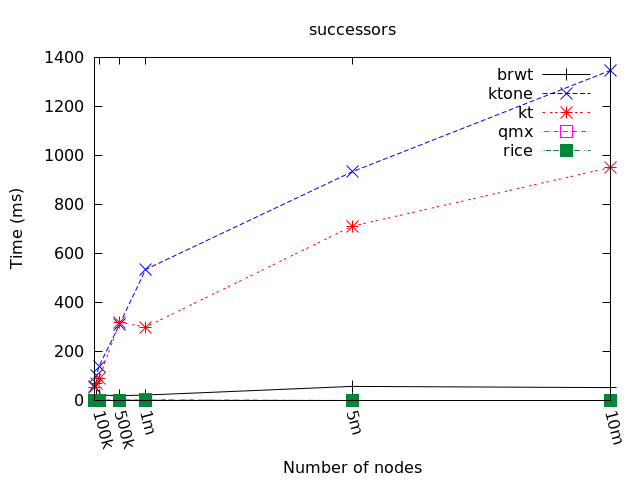}}
    \caption{successors}
    \end{subfigure}
    \begin{subfigure}[predecessors]{0.45\textwidth}{\includegraphics[width=.98\textwidth]{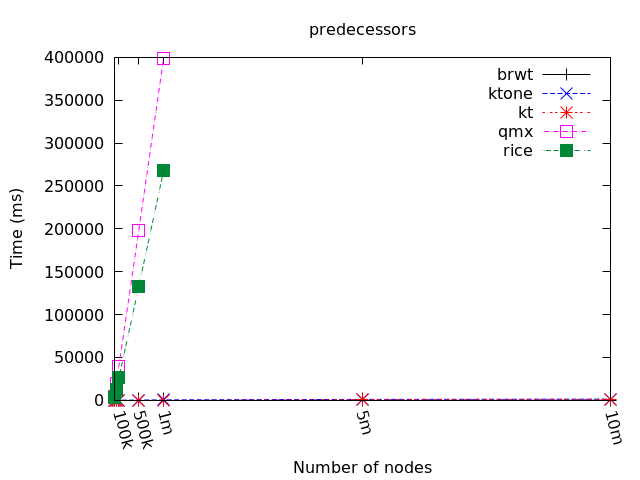}}
    \caption{predecessors}
    \end{subfigure}
    \begin{subfigure}[predecessors (compact)]{0.45\textwidth}{\includegraphics[width=.98\textwidth]{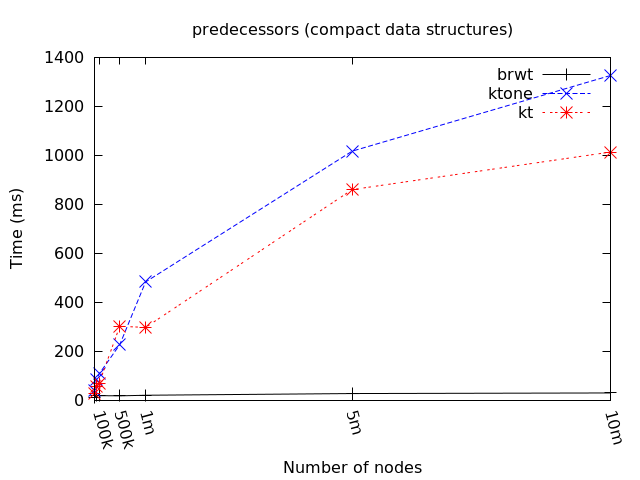}}
     \caption{predecessors (compact)}
     \label{fig8d}
    \end{subfigure}
    \caption{Number of nodes versus average time on neighborhood operations for scalability.}
    \label{fig:scala:search_ops}
\end{figure*}

Let us first analyze the neighborhood operations shown in Figure \ref{fig:scala:search_ops}. 
In general, we can see that the data structures behavior is as expected
from the previous analysis. The encoders scale very well for the isRelated-True and successors 
operations. Note that we have chosen the isRelated-True operation instead of isRelated, in order 
to force the structures to either navigate down the tree (compact data structures) or decompress
a non-empty list (for the encoders). We also concluded that the encoders performed badly for
predecessors, and this gets confirmed here. In fact, this operation could not be
completed for the relations with 5 and 10 million nodes, with our hardware configuration. If we
remove the encoders from the plot (Figure \ref{fig8d}), we can see that the BRWT scales well (almost constant)
while the remaining compact data structures scale in a logarithmic order. It is also worth
noticing that the $k^2$-tree and $k^2$-tree1 have a similar trend in all operations.

\begin{figure*}[t]
    \centering
    \begin{subfigure}[Escalating number of nodes]{0.45\textwidth}{\includegraphics[width=.98\textwidth]{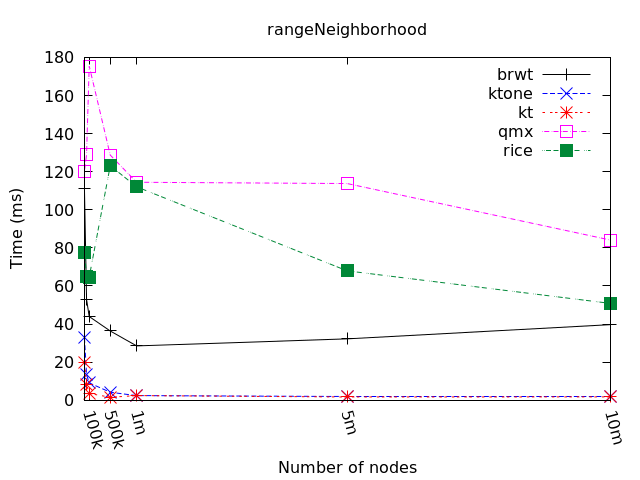}}
    \caption{Varying the number of nodes}
    \end{subfigure}
    \begin{subfigure}[Escalating range size]{0.45\textwidth}{\includegraphics[width=.98\textwidth]{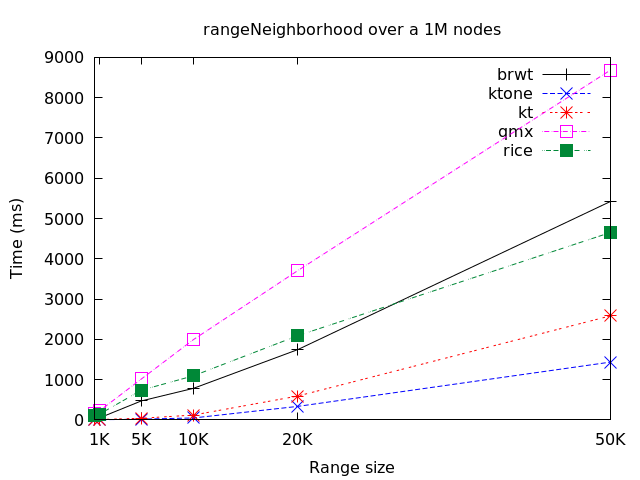}}
    \caption{Varying the range size}
    \end{subfigure}
    \caption{Scalability measures for rangeNeighborhood queries}
    \label{fig:scala:range}
\end{figure*}

For the \emph{rangeNeighborhood} operation, shown in Figure~\ref{fig:scala:range}, we have tested the scalability
in two different ways: varying the number of nodes (as in the previous cases) but maintaining a fixed
range size of $500\times 500$, and fixing the number of nodes but varying the range size (up to
$50,000\times 50,000$). It might seem strange that QMX and Rice-runs solve the range queries faster
when the number of nodes increases. However, this can be explained, as when the number of nodes increases,
the number of empty rows will probably increase too. Therefore, the number of rows actually explored and
decompressed decreases, so the time to solve the query is shorter. Note that the behavior of the
compact data structures is quite similar (it is not decreasing, but almost constant). 
Considering the variation of the range size, we can see that all data structures increase the time
for longer ranges, but in a sublinear order in general, being the  $k^2$-tree and $k^2$-tree1
the most efficient data structures.

\begin{figure*}[t]
    \centering
    \begin{subfigure}[Union]{0.45\textwidth}{\includegraphics[width=.98\textwidth]{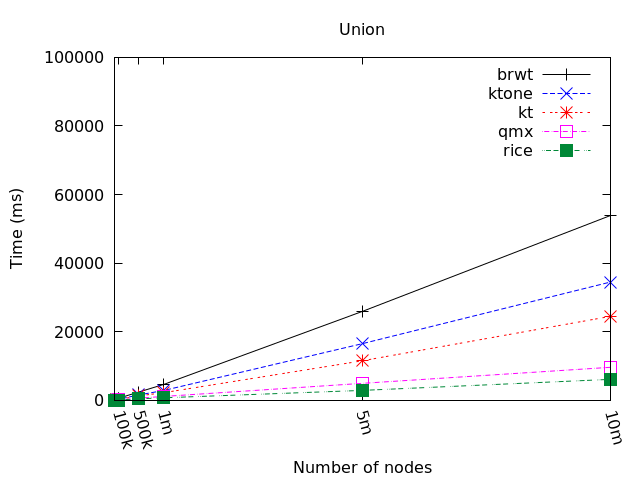}}
     \caption{Union}
    \end{subfigure}
    \begin{subfigure}[Intersection]{0.45\textwidth}{\includegraphics[width=.98\textwidth]{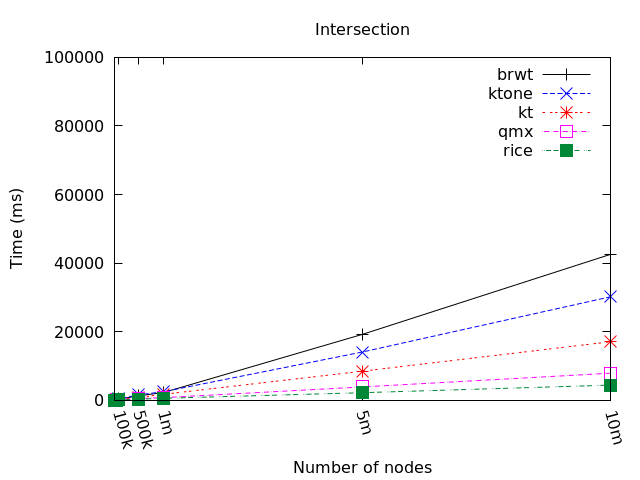}}
     \caption{Intersection}
    \end{subfigure}
    \begin{subfigure}[Difference]{0.45\textwidth}{\includegraphics[width=.98\textwidth]{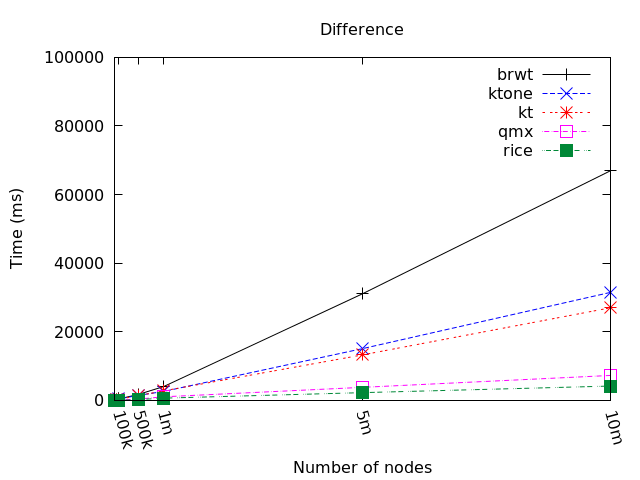}}
     \caption{Difference}
    \end{subfigure}
    \begin{subfigure}[Symmetric Difference]{0.45\textwidth}{\includegraphics[width=.98\textwidth]{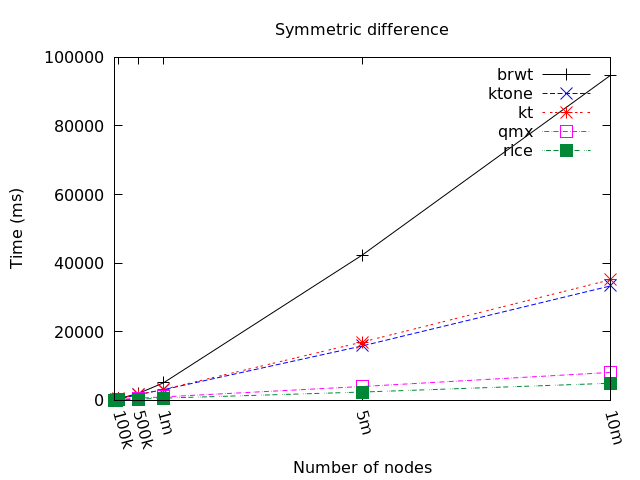}}
     \caption{Symmetric Difference}
    \end{subfigure}
    \caption{Number of nodes versus average time on set operations for scalability.}
    \label{fig:scala:set_ops}
\end{figure*}

For the set operations, illustrated in Figure~\ref{fig:scala:set_ops}, all data structures scale quite well, 
and in a uniform way (note that there are no crosses among the lines in the figure). We can highlight
that the BRWT is the worst option for these operations, while the encoders are the more suitable ones.
This confirms what was shown in Section~\ref{subsec:timings}.

\begin{figure}[t]
    \centering
    \includegraphics[width=.48\textwidth]{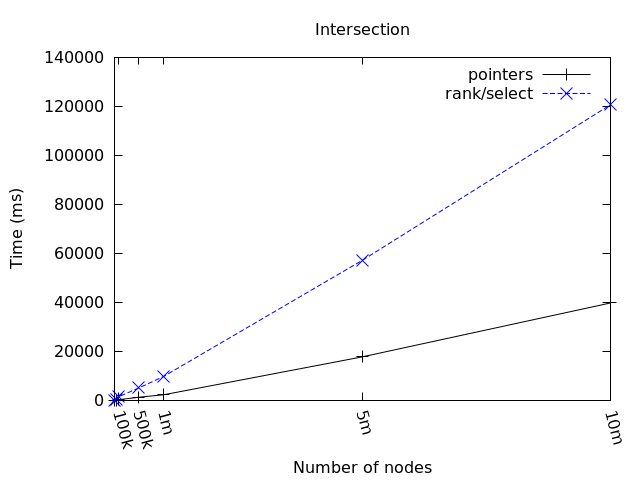}
    \caption{Speed improvement of queries on BRWT by using pointers}
    \label{fig:rank.vs.ptr}
\end{figure}

%reviewer 1 Q 4
A final note about scalability, but regarding some implementation
decisions for our algorithms: as we mentioned in Section~\ref{subsec-profundidad}, 
we decided to use a set of pointers instead of using the \texttt{rank} and \texttt{select}
operations to speed up the depth-first set operations over BRWTs. The same decision
was taken to speed up the operations over both variants of $k^2$-trees.
Figure~\ref{fig:rank.vs.ptr} shows the behavior of the intersection algorithm using both implementations.
The version with pointers clearly outperforms the \texttt{rank}/\texttt{select}
version, especially for large datasets (up to 3 times faster). Of course, this speed-up comes
with a price, because the pointer version takes up more memory (between 30\% and 56\%).

\section{Conclusions}\label{sec:conclusions}
In this work, we have conducted several experiments to compare the behavior of several
data structures used to store binary relations. We have considered three compact data
structures ($k^2$-tree, $k^2$-tree1 and BRWT) and two encoders or compressors (QMX and
Rice-runs). 

For the compact data structures, we used the algorithms for $k^2$-trees and
$k^2$-tree1s developed in \cite{BLN14} and \cite{QUIJADAFUENTES201976}, but the algorithms
for set operation over the BRWT are presented here for the first time, thus extending the
functionality of this data structure. 

We have found that there is no clear winner, no data structure is better than the
rest in all cases. All of them have some advantages and disadvantages, depending on
several factors. We have considered the storage size and the response time as basic
measurements, and have tested them using several datasets with different characteristics,
because the data distribution has a great impact on the performance of all data structures.

In order to offer some general conclusions, we can group the data structures in three groups 
that have a similar behavior: the encoders (QMX and Rice-runs), both $k^2$-tree variants, and BRWT.

With respect to the encoders, they proved to be the fastest option for the set operations in all
cases, and are competitive for some neighborhood queries, except for \emph{rangeNeighborhood}
and especially the \emph{predecessors} queries (which could not actually be executed for 
large datasets). In terms of storage needs, the encoders use in general more space than the compact
data structures.

The $k^2$-trees excel at the \emph{rangeNeighborhood} queries, but are outperformed for
the \emph{successors} operation. For the rest of the neighborhood queries they are competitive.
For the set operations, these structures are not as fast as the encoders, but they are the
best option amongst the compact data structures. They also scale reasonably well when the
dataset grows. In terms of storage, $k^2$-trees are always the best option, using much less 
space than the other structures. This can let the $k^2$-tree be a good option for those
operations where they are slower than the encoders, when the encoders cannot fit the relation
into main memory.

BRWT is competitive for the neighborhood queries in general, and is the best option for 
the \emph{predecessors} queries. For the set operations, however, it is usually the worst
option. In terms of storage needs, it is competitive with respect to the remaining compact
data structures, and better than the encoders.

Finally, we must highlight that the data distribution has a great impact on both the
size of the data structure and the speed of the operations. In general, clustered
data distributions tend to favor compact data structures, while more random or evenly
distributed datasets tend to benefit the encoders.

We have presented here, to the best of our knowledge, the first study about the behavior
of compressed data structures for binary relations, evaluating storage needs and speed of the operations based
on different (synthetic and real) data distributions, considering also the scalability
of the data structures.

\bibliographystyle{IEEEtran}
\bibliography{refs}

\clearpage
\begin{IEEEbiography}[{\includegraphics[width=1in,height=1.25in,clip,keepaspectratio]{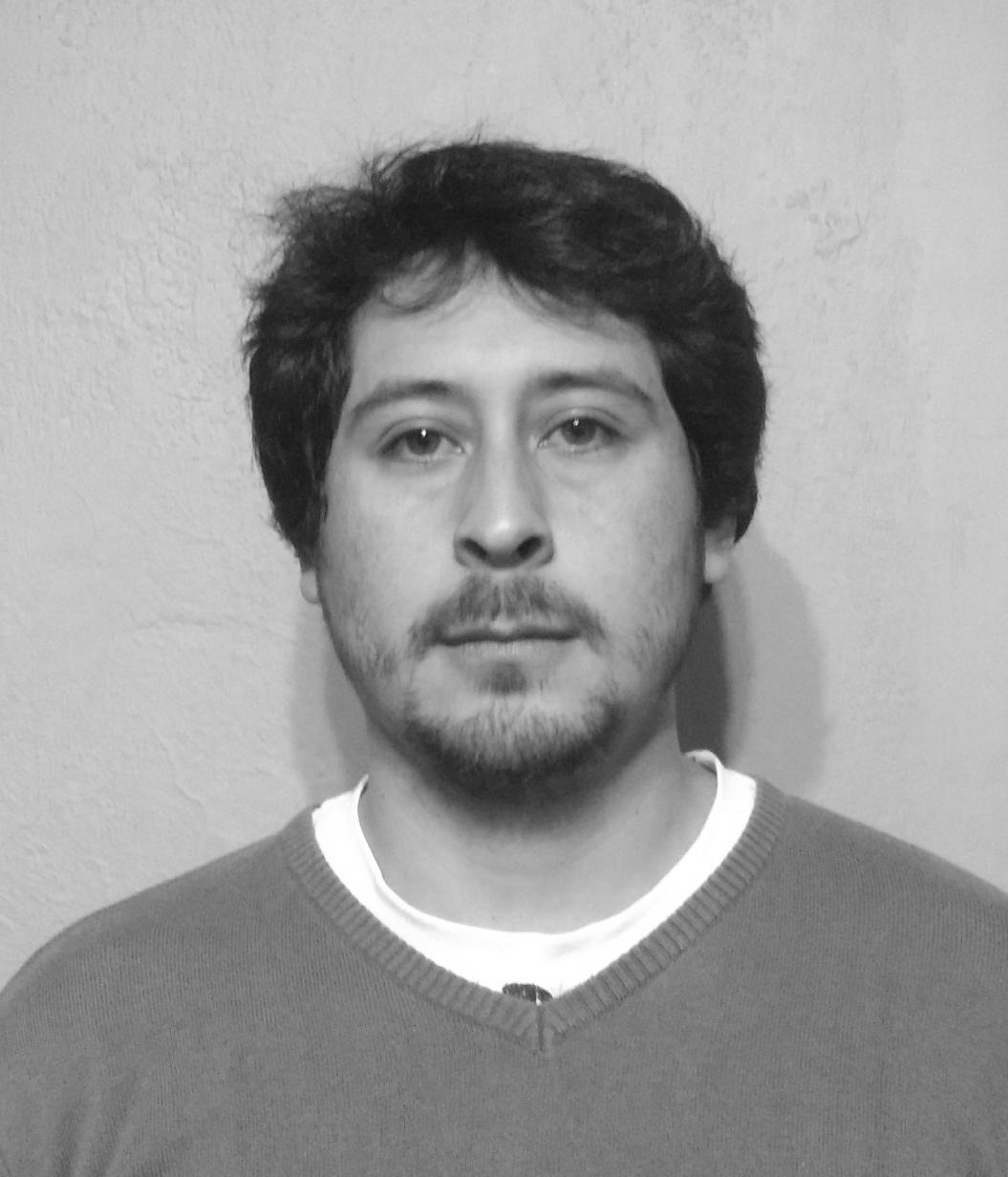}}]{Carlos Quijada Fuentes} obtained his degree in Civil Engineering in Computer Science in 2011
and his Master in Computer Science in 2017, both from the University of B\'io-B\'io. His research area 
is data structures and algorithms. Chill\'an/Chile.
\end{IEEEbiography}

\begin{IEEEbiography}[{\includegraphics[width=1in,height=1.25in,clip,keepaspectratio]{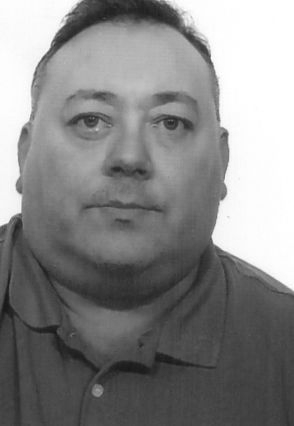}}]{Miguel R. Penabad} obtained his Master in Computer Science in 1994 at the University of A Coru\~na.  He received his Ph.D in 2001 at the University of A Coru\~na. He is a professor in the same university since 2000. His main research interests are database query optimization, and algorithms and data structures for information retrieval.
\end{IEEEbiography}

\begin{IEEEbiography}[{\includegraphics[width=1in,height=1.25in,clip,keepaspectratio]{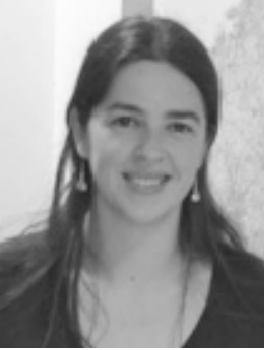}}]{Susana Ladra} received the bachelor's degree
in mathematics from the National Distance Education
University (UNED), in 2014, and the master's
in computer science engineering and the Ph.D.
degree in computer science from the University of
A Coruña, in 2007 and 2011, respectively. She is
currently an Associate Professor with the Universidade
da Coruña. She is the Principal Investigator
of several national and international research
projects. She has published more than 40 articles in
various international journals and conferences. Her research interests include
design and analysis of algorithms and data structures, and data compression
and data mining in the fields of information retrieval and bioinformatics.
\end{IEEEbiography}

\begin{IEEEbiography}[{\includegraphics[width=1in,height=1.25in,clip,keepaspectratio]{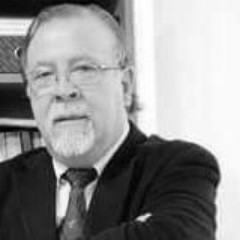}}]{Gilberto Guti\'errez Retamal}
received his M. Sc. from the University of Chile in 1999 and the Ph.D. in computer science in 2007 from the same university. His research areas include data structures and algorithms, spatial and spatio-temporal databases.
He is currently an associate professor in the Department of Computer Science and Information 
Technology, at the University of B\'io-B\'io, Chill\'an / Chile.
\end{IEEEbiography}

\EOD

\end{document}